\documentclass[twocolumn]{aa}  
\usepackage{graphicx}
\usepackage[varg]{txfonts}
\usepackage{xspace}
\usepackage{amsmath}
\usepackage{placeins}
\usepackage{gensymb}
\usepackage{color}
\usepackage{float}
\usepackage{todonotes}

\renewcommand{\vec}[1]{\mbox{\boldmath$#1$}}
\newcommand{\music}{\texttt{MUSIC}\xspace}

\definecolor{orange}{rgb}{.9,.3,0}
\definecolor{mypink1}{rgb}{0.858, 0.188, 0.478}

\begin{document}
\title{The shape of convection in 2D and 3D global simulations of stellar interiors}
\author{M.-G. Dethero\inst{\ref{inst2}} \and J. Pratt\inst{\ref{inst1},\ref{inst2},\ref{inst3}} \and D.G. Vlaykov\inst{\ref{inst3}}  \and I. Baraffe\inst{\ref{inst3},\ref{inst4}} 
\and T. Guillet\inst{\ref{inst3}} \and T. Goffrey\inst{\ref{inst5}}
\and A. Le Saux \inst{\ref{inst7}}
\and A. Morison\inst{\ref{inst3}}
 }
\institute{Lawrence Livermore National Laboratory, 7000 East Ave, Livermore, CA 94550, USA \email{pratt34@llnl.gov}\label{inst1}
\and Department of Physics and Astronomy, Georgia State University, Atlanta GA 30303, USA \label{inst2} 
\and Astrophysics, College of Engineering, Mathematics and Physical Sciences, University of Exeter, EX4 4QL Exeter, United Kingdom \label{inst3}
\and \'Ecole Normale Sup\'erieure de Lyon, CRAL (UMR CNRS 5574), Universit\'e de Lyon 1, 69007 Lyon, France\label{inst4}
\and Centre for Fusion, Space and Astrophysics, Department of Physics, University of Warwick, Coventry, CV4 7AL, United Kingdom \label{inst5}
\and  Laboratoire de Meteorologie Dynamique (IPSL), Sorbonne University, CNRS, Ecole Polytechnique, Ecole Normale Superieure, Paris, France \label{inst7}
} 

\titlerunning{The shape of convection in stellar interiors}
\authorrunning{M.-G. Dethero et. al.}

\abstract
 {Theoretical descriptions of convective overshooting in stellar interiors often rely on a basic one-dimensional parameterization of the flow called the filling factor for convection.  Several different definitions of the filling factor have been developed for this purpose, based on: (1) the percentage of the volume, (2) the mass flux, and (3) the convective flux that moves through the boundary.}
{We examine these definitions of the filling factor with the goal of establishing their ability to explain differences between 2D and 3D global simulations of stellar interiors that include fully compressible hydrodynamics and realistic microphysics for stars.}
{We study convection and overshooting in pairs of identical two- and three-dimensional global simulations of stars produced with \music, a fully compressible, time-implicit hydrodynamics code.  We examine pairs of simulations for  (1) a $3$ M$_{\odot}$ red giant star near the first dredge-up point, (2) a $1$ M$_{\odot}$ pre-main-sequence star with a large convection zone, (3)  the current sun, and (4) a $20$ M$_{\odot}$ main-sequence star with a large convective core.}
{Our calculations of the filling factor based on the volume percentage and the mass flux indicate asymmetrical convection near the surface for each star with an outer convection zone.  However, near the convective boundary, convective flows achieve inward-outward symmetry for each star that we study; for 2D and 3D simulations, these filling factors are indistinguishable.   A filling factor based on the convective flux is contaminated by boundary-layer-like flows, making a theoretical interpretation difficult.  We present two possible new alternatives to these frequently used definitions of a filling factor, which instead compare flows at two different radial points.  The first alternative is the \emph{penetration parameter} of \citet{anders2022stellar}. The second alternative is a new statistic that we call the \emph{plume interaction parameter}.  We demonstrate that both of these parameters captures systematic differences between 2D and 3D simulations around the convective boundary.
}
{}
\maketitle



\section{Introduction \label{sec:intro} }

Predicting the evolution of stars requires a prescription for the amount of mixing caused by convection at different values of the internal radius of the star.  This mixing has been widely linked to the concept of a filling factor for the convective inflows \citep[e.g.][]{schmitt1984overshoot,stein1989topology,cattaneo1989two,toomre1990three,cattaneo1991turbulent,zahn1991convective,canuto1998stellar,brummell2002penetration,rempel2004overshoot,browning2004simulations,rogers2006numerical,kapyla2017extended,cai2020upward}.  Conceptually, a filling factor quantifies, based on fluid motion or heat, how much of the star is moving inward at a given radius.  
Several early works on stellar convection \citep[e.g.][]{schmitt1984overshoot,stein1989topology,hurlburt1984two} predict that the large stratification present in stellar convection zones should produce convective motions that have a pronounced asymmetry between inflows and outflows, corresponding to a filling factor less than one-half.   Indeed observations of the solar surface find 
strong, coherent inflows surrounded by weaker, more diffuse outflows \citep[e.g. as discussed in][]{nordlund2009solar}. 
Where present, such an asymmetry would alter the shear interaction between opposing flow structures, and thus change the amount of fluid mixing.   The filling factor has thus been viewed as a fundamental measure of how the stellar convection zone is structured.

Evaluating the filling factor at the boundary of convective instability provides a measure of how much fluid overshoots the convection zone and enters the radiative zone. This provides a link between the structure of convective flows in the convection zone and the depth of convective overshooting or penetration. 
Several works \citep{zahn1991convective,brummell2002penetration} suggest that a filling factor could play a more significant role in determining the overshooting depth than the radial velocity at the convective boundary. The convective velocities typically differ between 2D and 3D, and, as noted in these works, the convective flow structures are visually different;
the filling factor has been discussed as a plausible source of difference between two-dimensional and three-dimensional stellar simulations \citep[e.g.][]{schmitt1984overshoot,cai2020upward,pratt2020comparison}.

The question of how to best define a filling factor remains open because it depends on how to accurately define the edges of structures in a convective flow. Without loss of generality, we call such structures ``plumes'' in this work.  We will also refer to convective flows moving inward toward the core as ``inflows'' and those moving outward toward the surface as ``outflows.''  Many works \citep[e.g. ][]{schmitt1984overshoot,canuto1998stellar,brummell2002penetration,andrassythesis}
have defined the filling factor to be the fractional area occupied by inflows at a given radius interior to the star.  These works thus define the edges of convective plumes using the radial velocity.  We call this definition the \emph{volume-percentage} filling factor.  In contrast, \citet{rempel2004overshoot} defines a filling factor as the fraction of the volume at a given radius with an inward mass flux; we call this the \emph{mass-flux} filling factor. \citet{zahn1991convective} defines a plume based on the convective flux, and uses the fraction of the convective flux carried by inflows to define a filling factor; we call this the \emph{convective-flux} filling factor.  In a similar vein, \citet{anders2022stellar} propose a ``penetration parameter,'' based on the change in the convective flux between the convection zone and the overshooting zone, as a predictive one-dimensional quantity.  

These definitions of the filling factor have been evaluated in ideal box-in-a-star type simulations, which often use a moderate stratification and an ideal gas equation of state. However, they have not been examined in the kinds of global stellar simulations that we perform in this work.  Our simulations are performed in a spherical shell that includes a large portion of the stellar radius and uses a stratification, temperature gradient, equation of state, and opacity tables that are extracted directly from stellar structure models accurate to the current state-of-the-art in stellar modeling.  Our simulations also solve the equations for fully compressible convection; no additional assumptions are made that could impact the asymmetry of the convective flow.\footnote{We will generally refer to this combination of choices for modeling stellar interiors using a stratification and microphysics directly from stellar structure and evolution calculations, a spherical volume, and compressible hydrodynamic equations as \emph{realistic}, as opposed to more idealized models.}  We evaluate each definition of a filling factor for suitability to use in theoretical models of convection and overshooting based on two criteria, namely whether they: (1) are different for different stellar models and for 2D and 3D simulations, and (2) are correlated with the measured overshooting depth in the simulations.
 
This work is structured as follows. In Section \ref{sec:sim}, we will discuss the global fluid simulations of stars that we use to study the filling factor.  In Section \ref{sec2d3dbench} we describe observed differences between 2D and 3D simulations of stars.  In Section \ref{secresultsid} we describe how the volume-percentage, mass-flux, and convective-flux filling factors are calculated and present results from our simulations. We also evaluate calculations of the Anders penetration parameter. In Section \ref{secnewstuff3} we present statistics based on the width and numbers of inflows and outflows.  We use these diagnostics to build a new nondimensional parameter, which we call the plume interaction parameter, and we show that this parameter is systematically different in 2D and 3D simulations for the stars we examine.   In Section \ref{secconc}, we discuss the implications of our results for future stellar evolution calculations.

\section{Simulations \label{sec:sim}}

\begin{figure*}
\begin{center}
\includegraphics[height=3in]{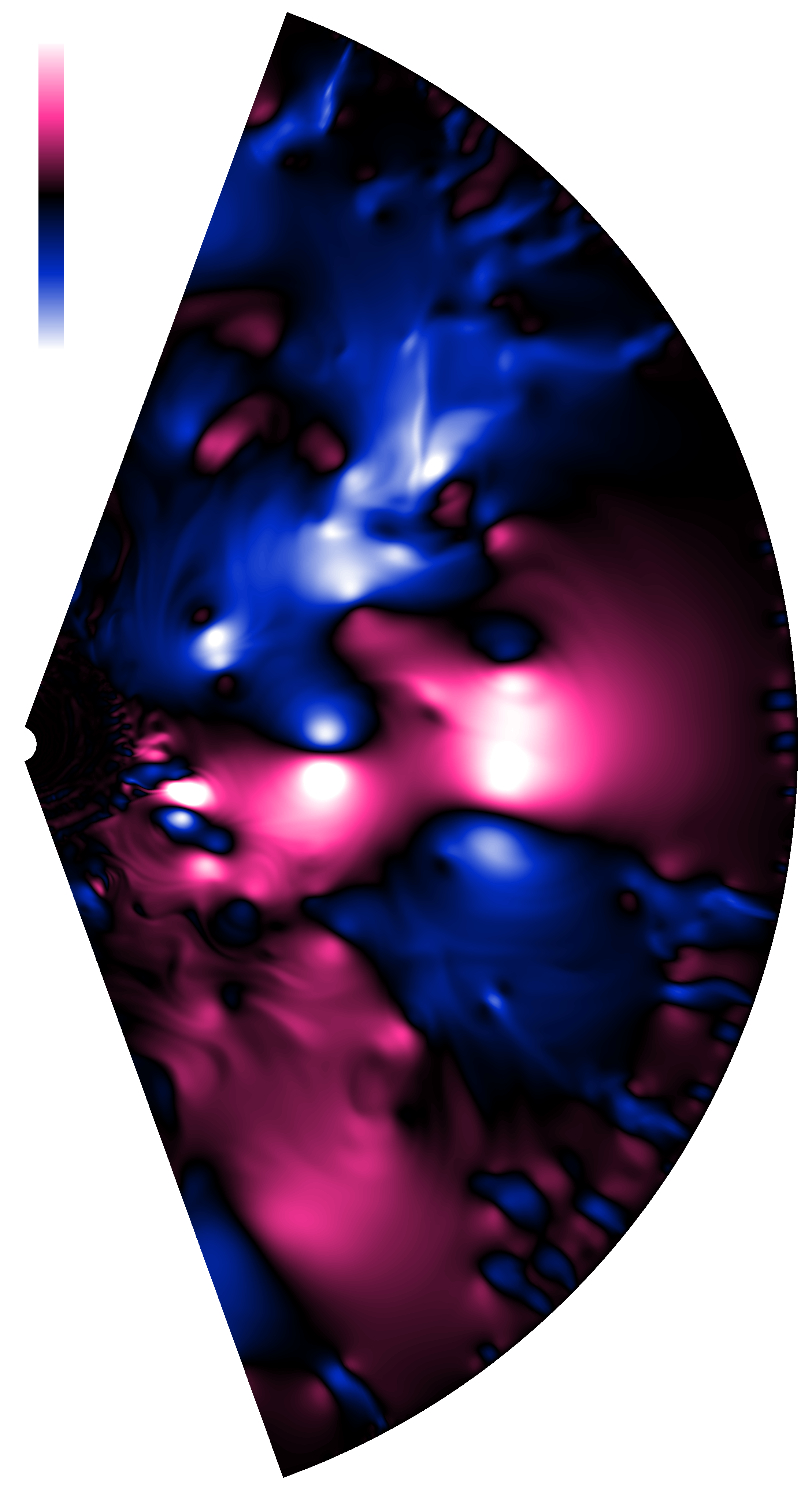}\includegraphics[height=3in]{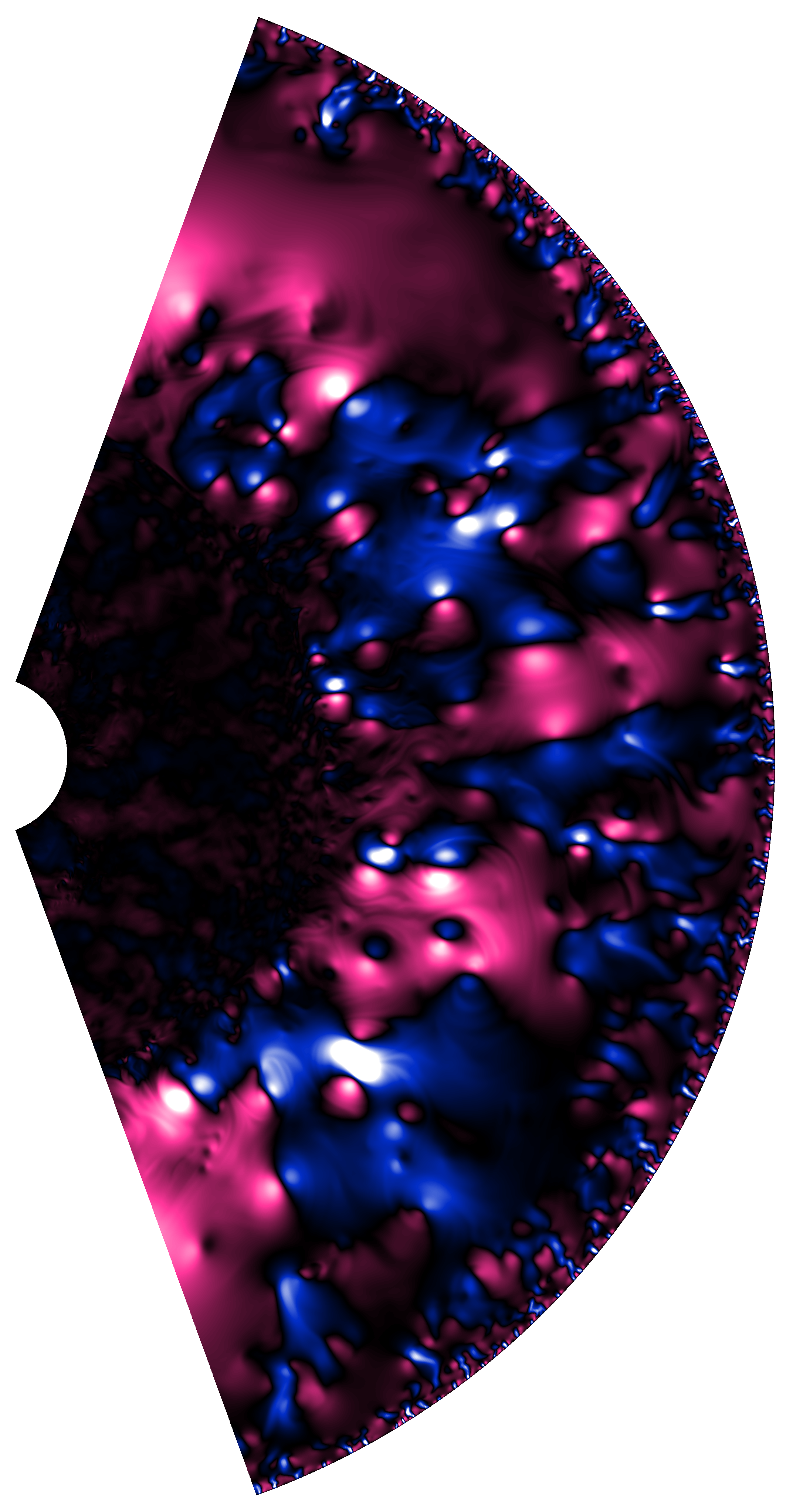}\includegraphics[height=3in]{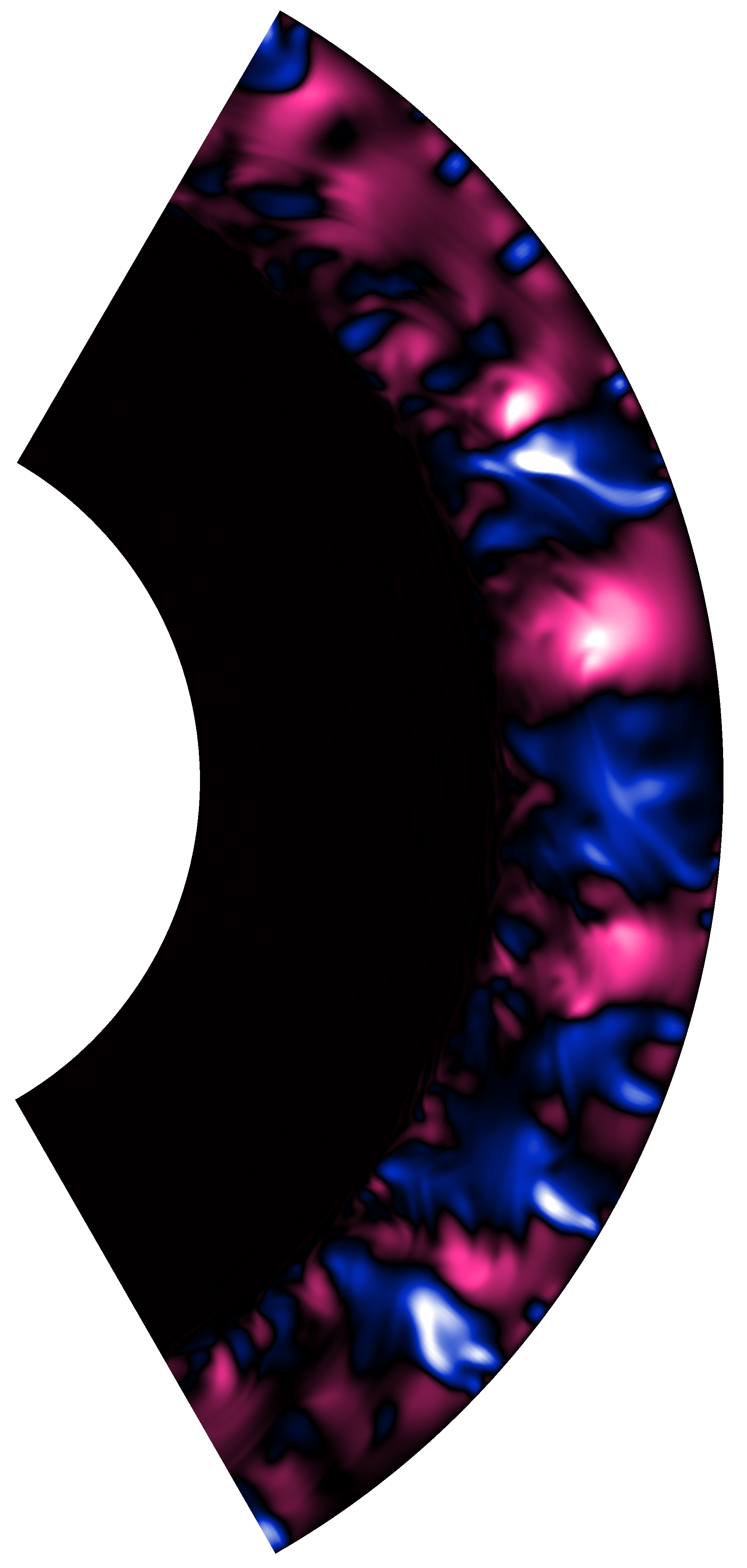}\includegraphics[height=3in]{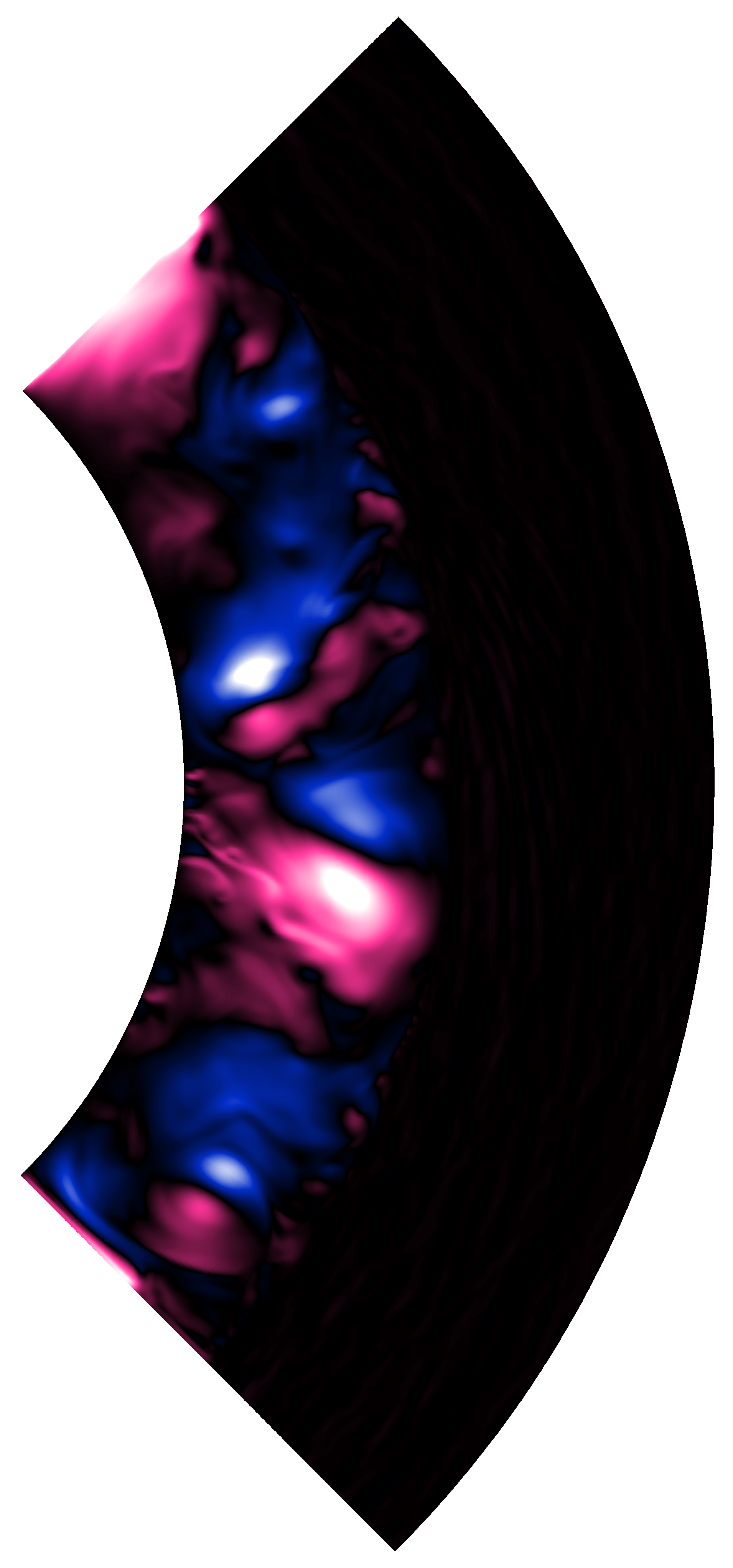}
\caption{Visualizations of radial velocity in 2D simulations (from left to right) \emph{bg2D}, \emph{wm2D}, \emph{dcs2D}, and \emph{cc2D}.  Outward flows are indicated in pink, while inward flows are in blue; the zero point in velocity is black. {The maximum and minimum values of the color scale are defined by a radial velocity magnitude near the maximum for each simulation: \emph{bg2D} ($\pm 6.5$ km/s), \emph{wm2D} ($\pm 2.9$ km/s), \emph{dcs2D} $ (\pm 0.63$ km/s), \emph{cc2D} ($\pm 1.5$ km/s).}
\label{allviz}}
\end{center}
\end{figure*}

We produce simulations of stellar convection in four different stellar structures.  
One star that we examine is a $3$ M$_\odot$ star that is ascending the red giant branch, produced with the open-source Modules for Experiments in Stellar Astrophysics (MESA) code \citep{paxton2010modules}; this star has a large outer convection zone because it is near the first dredge-up point.
The second star that we examine is a one M$_{\odot}$ pre-main sequence star, called the young sun, produced with the Lyon stellar evolution code \citep{baraffe1991evolution, baraffe1997evolutionary,baraffe1998evolutionary}.  This model has been described and examined in \citet{pratt2016spherical,pratt2017extreme,pratt2020comparison}; we refer to that earlier work for additional details of the young sun, beyond the brief summary of relevant points given here.     The third star is a one M$_\odot$ main-sequence star with a moderate outer convective layer, similar to the current sun, produced with the Lyon stellar evolution code.  The fourth star is a $20$ M$_\odot$ star at the zero-age main sequence (ZAMS) with a large convective core, produced with the Lyon stellar evolution code. These stars represent the different stages of evolution where convection occurs: pre-main sequence, main sequence, and evolved stars.  They also represent different types of convection zones: outer convective envelopes that range between deep and shallow, as well as convective cores.  Visualizations of the radial velocities in each of these stars are provided in Fig.~\ref{allviz}.

In these simulations, each star is assumed to have a homogeneous chemical composition.  We perform pairs of 2D and 3D Implicit Large Eddy Simulations (ILES) \citep{grinstein2007implicit,ritos2018performance,margolin2019reality} of these stars using the {MUlti-dimensional} Stellar Implicit Code (\music).   In these pairs, the simulation volume and grid for the 3D simulation are identical to the 2D simulation except for the dimensionality,  as in \citet{pratt2020comparison}.  Our simulations in this work only take convection into account; the possibility of studying additional physical effects such as rotation, a tachocline, chemical mixing, and magnetic fields is omitted from the current study.

We use the \music code to solve the inviscid compressible hydrodynamic equations for density $\rho$, momentum $\rho \vec{u}$, and internal energy $\rho e$:
\begin{eqnarray} \label{densityeq}
\frac{\partial}{\partial t} \rho &=& -\nabla \cdot (\rho \vec{u})~,
\\ \label{momeq}
\frac{\partial}{\partial t} (\rho \vec{u}) &=& -\nabla \cdot (\rho \vec{u} \otimes \vec{u}) - \nabla p + \rho \vec{g} ~,
\\ \label{ieneq}
\frac{\partial}{\partial t} (\rho e) &=& -\nabla \cdot (\rho e\vec{u}) - p \nabla \cdot \vec{u} + \nabla \cdot (\chi \nabla T) ~,
\end{eqnarray}
using a second-order finite volume method, a MUSCL method \citep{van1997towards, thornber2008improved} of interpolation on a staggered grid, and a van Leer flux limiter \citep[as described in][]{van1974towards,roe1986characteristic,leveque2006computational}.   
For 2D simulations, the finite volume method assumes azimuthal symmetry.  Time integration in the \music code is fully implicit and uses a Jacobian free Newton-Krylov (JFNK) solver \citep{knoll2004jacobian} with physics-based preconditioning \citep{maximepaper,newman2013physics,mousseau2000physics,chen2014fluid,holod2021enhanced}.  The \music code uses an adaptive time step, which is constrained identically for each pair of 2D and 3D simulations.  \music simulations use the same tabulated equation of state and opacity that are used by the 1D stellar evolution code that produced the stellar structure.  In Eq. \eqref{momeq}, $\vec{g}$ is the gravitational acceleration, a spherically symmetric vector consistent with that used in the stellar evolution calculation.  It is not evolved for any of the simulations with an outer convective envelope; the convective core simulation did allow the gravity to be recalculated, but changes in this gravity term were small (see \citet{baraffe2023study}).

\begin{table*}
\begin{center}
\caption{Parameters for compressible hydrodynamic simulations performed with \music.  Two different pairs of young sun simulations are included: (1) the \emph{wide2D} and \emph{wide3D} pair have lower radial resolution but the 3D simulation uses a large angle along the equator, while (2) \emph{wm2D} and \emph{wm3D} use a smaller equatorial angle but significantly higher radial resolution.  The pair of red giant simulations is \emph{bg2D} and \emph{bg3D}; the pair of current sun simulations is \emph{dcs2D} and \emph{dcs3D}; the pair of convective core simulations is \emph{cc2D} and \emph{cc3D}.    The simulation name, dimensionality, evolutionary state, and stellar mass $M$ in units of the solar mass $M_{\odot}$ are provided.  The inner radius $R_{\mathsf{i}}$ of the spherical shell, the radius of the convective boundary $R_{\mathsf{CB}}$ {determined by the Schwarzschild criterion}, and the outer radius $R_{\mathsf{o}}$ of the spherical shell for the simulation are given in units of the total radius of each star, $R$, as a triplet.   The angular extent of the simulation in the polar and equatorial directions is given as $(\theta,\phi)$, and the grid spacing in both angular directions is $(\Delta \theta,\Delta \phi)$.  The average global convective turnover time $\tau_{\mathsf{conv}}$ is provided as well as its standard deviation in time, and the total time span for each simulation is given in units of the convective turnover time.
 \label{simsuma}
 }
\begin{tabular}{lcccccccccccccccccccccccccccc}
                      & dims. &  evol. &  $M/M_{\odot}$ &  $(R_{\mathsf{i}},R_{\mathsf{CB}},R_{\mathsf{o}})/R$  & $(\theta,\phi)$ (\degree) & $(\Delta \theta,\Delta \phi)$ (\degree)  &  $H_{p,\mathsf{CB}}/\Delta r$ & $\tau_{\mathsf{conv}}$($10^5$s) & time ($\tau_{\mathsf{conv}}$) 
\\ \hline
wide3D               & 3D & PMS &  1  & (0.21,0.43, 1.00)  &   (140, 140) & (0.55,0.55) & 63    &  $ 7.06 \pm 0.3 $  & 3.8  
\\ \hline
wide2D               & 2D & PMS & 1 & (0.21,0.43, 1.00)   &   (140,0) & (0.55,-) &  63   &  $7.9 \pm 0.7 $  & 104 
\\ \hline
wm3D                & 3D & PMS & 1 & (0.10,0.43, 1.00)   &  (140,9) & (0.14,0.14) & 258    &  $ 3.91 \pm 0.21 $  & 5.4  
\\ \hline
wm2D                & 2D & PMS & 1 & (0.10,0.43, 1.00)  &    (140,0)  & (0.14,-) &  258   &  $ 3.76 \pm 0.98 $  & 141      
\\ \hline
bg3D                & 3D & RGB &  3 & (0.02, 0.175,0.90)  & (140,19) & (0.30,0.30) & 55    &  $ 6.48 \pm 0.07 $  & 22  
\\ \hline
bg2D                & 2D & RGB & 3 & (0.02, 0.175,0.90)  &   (140,0) & (0.30,-) &  55   &  $ 10.86 \pm 2.16 $  & 163     
\\ \hline
cc3D                & 3D & MS & 20 & (0.194, 0.287, 0.38) &  (90,22) & (0.27,0.52) & 140    &  $ 92.6 \pm 7.7 $  & 3.77  
\\ \hline
cc2D                & 2D & MS & 20 & (0.194, 0.287, 0.38)  &  (90,0)  & (0.27,-) &  140   &  $ 36.5 \pm 27.5  $  & 35.7      %
\\ \hline
dcs3D                & 3D & MS & 1 &  (0.40, 0.72, 0.97)   &   (120,240) & (0.33, 0.64) & 134    &  $ 9.6 \pm 1.0 $  & 3.9  
\\ \hline
dcs2D                & 2D & MS & 1 & (0.40, 0.72, 0.97)   &   (120,0) & (0.33,-) &  134   & $28.4 \pm 5.5$ & 84 %
\\ \hline
\end{tabular}
\end{center}
\end{table*}

We study the pairs of \music simulations in 2D and 3D described in Table~\ref{simsuma}.  
In all of our \music simulations, the compressible hydrodynamic equations \eqref{densityeq}-\eqref{ieneq} are solved in a spherical shell using spherical coordinates:  radius $r$, and angular variables $\theta$ and $\phi$ (in 3D).  
As we noted in \citet{pratt2020comparison}, grid spacing is particularly important in determining the physics in ILES, because it is directly related to the effective numerical viscosity. 
 In Table~\ref{simsuma}, the inner and outer radius of the spherical shell for each simulation is noted, and the radial and angular grid spacings are specified.
 The simulation volume and grid in the $r$ -- $\theta$ plane are identical for each pair of simulations. {To obtain accurate statistics on overshooting, both radial resolution and a long-time series of data are critical. For that reason, it is ideal for us to examine convection in spherical shells. \citet{herwig20233d} have recently described a dipole mode that fills the entire convective core in their simulation of a $25 M_\odot$ main-sequence star; that mode is clearly not possible in either of the simulations of the main-sequence star that we examine here.  Nevertheless, the 2D and 3D simulations of this star both equally neglect the possibility of such a dipole mode dominating the flow.   We leave further study of dipole modes in convective cores to future work.}
 
To allow a clear comparison of the convective turnover time between 2D and 3D simulations, we define this fundamental parameter as in \citet{pratt2016spherical}:
\begin{eqnarray}\label{eqtauconv}
\tau_{\mathrm{conv}}(t) =  \int_{\mathsf{CZ}} d \mathsf{V}~ \frac{H_p}{|v|}  ~\Big/~ \int_{\mathsf{CZ}}  d \mathsf{V}
\end{eqnarray}
In this expression, $H_p$ is the pressure scale height, and the magnitude of velocity in the denominator is calculated in either 2D or 3D, depending on the simulation. The integration covers the convection zone (CZ) and is volume-weighted using volume element $d\mathsf{V}$.  The instantaneous value of $\tau_{\mathrm{conv}}$ in eq.~\eqref{eqtauconv} is averaged over time to produce the values in Table \ref{simsuma}.
 
In Table \ref{simsuma}, we introduce the quantity $H_{p,\mathsf{CB}}/\Delta r$.  This ratio shows how many grid spaces resolve each pressure scale height at the convective boundary. {In this work, we will often refer to the convective boundary; this is the boundary of the region of convective instability calculated by the Schwarzschild criterion, which does not evolve during the simulations in Table~\ref{simsuma}.}  Our highest resolution simulation pair is \emph{wm2D/wm3D}. Simulation \emph{wm3D} has a grid of $r \times \theta \times \phi = 1312 \times 1024 \times 64$.  Our simulations have sufficient radial resolution to produce a characteristic radial profile for velocity in 2D. 

{All simulations in this work are ILES, and convergence is not expected in the absolute sense that direct numerical simulations (DNS) converge. A universal shape of the velocity profiles can be observed with sufficiently small grid spacing, and the increase in the velocities becomes less as the grid spacing is progressively decreased \citep[see also the discussion of ILES and convergence in ][]{Andrassy2024}.  A study of the effect of grid spacing was examined systematically for the young sun in \citet{pratt2016spherical}, and for the current sun in \citet{vlaykov2022impact}.  The main sequence core convection simulation was studied in \citet{baraffe2023study}.  Similar results are obtained for the red giant simulations.  
Because of the complications presented by ILES for convergence, we find the use of resolution criteria based on $H_{p,\mathsf{CB}}/\Delta
r$ to be more useful for convective overshooting than the traditional DNS-style convergence studies.   Such resolution criteria allow for clear comparisons between simulations of different stars that use different grids.
}

All data studied here are produced during steady-state convection, a period where the time-averaged value of the total kinetic energy is well-defined and not evolving in time.  Each 3D simulation includes at least 3 convective turnover times of data taken during steady-state convection; for each 2D simulation, the time span is more than 30 $\tau_{\mathsf{conv}}$. 
The uncertainties in the calculation of the average convective turnover time have a larger standard deviation for the 2D simulation than the 3D simulation; this is clearly impacted by the longer time series of data available for the 2D simulations \citep[e.g. also observed in ][]{pratt2020comparison}.

We examine simulations with two variations on energy boundary conditions that maintain the original radial profiles of density and temperature of the 1D stellar evolution model.  For the \emph{wide2D/wide3D} and \emph{wm2D/wm3D}  simulations, which include the full stellar radius up to the photosphere, the surface radiates energy with the local surface temperature.  In this case, the energy flux varies as $\sigma T_{\mathsf{s}}^4$, where $\sigma$ is the Stefan-Boltzmann constant and $T_{\mathsf{s}}(\theta,t)$ is the temperature along the surface.  This boundary condition can only be effectively used when the near-surface layers are included in the simulation volume and the temperature gradient near the surface is sufficiently resolved; otherwise, it results in artificially high cooling rates.  The young sun simulations have more than one grid space per pressure scale height in this region.  For the \emph{bg2D/bg3D}, \emph{cc2D/cc3D}, and \emph{dcs2D/dcs3D} simulations, which do not include the full stellar radius, we hold the energy flux and luminosity constant on the outer radial boundary, at values established by the stellar structure.  For an examination of how these boundary conditions affect the dynamics, we refer to \citet{pratt2016spherical} and \citet{vlaykov2022impact}.

{In all of the simulations studied in this work, we use the luminosity profiles accurate to the stellar structure models; we do not employ the tactic of luminosity boosting to shorten the convective turnover time and bring the thermal time-scale of the star closer to the convective turnover time.
Luminosity boosting leads to a substantial reduction in computational costs and makes reaching thermal equilibrium feasible when a large enough boost factor is used. However, it can also distort the original background stratification of the star. Even if this is avoided, as discussed in
\citet{baraffe2021two}, luminosity boosting increases the overshooting depth, the local heating in
the overshooting layer, and the shape of the spectrum of waves excited \citep{lecoanet2019low,le2022two}.  Our choice to use luminosities that have not been artificially boosted is motivated by the need to  measure an overshooting depth for a specific star.  The simulations that we study in this work are far from thermal equilibrium.  However the temperature gradient is not evolving during the course of these simulations, and the statistics that we produce are meaningful.}

For the velocity, we impose non-penetrative and stress-free boundary conditions in the radial directions for all simulations.  The energy flux and luminosity are held constant at the inner radial boundary, at the value of the energy flux at that radius in the one-dimensional stellar evolution calculation.
On the inner radial boundary of the spherical shell, we impose a constant radial derivative on the density for all simulations.  At the outer radial boundary, we apply different boundary conditions that suit the local derivatives in density best. For simulations \emph{bg2D/bg3D}, \emph{wide2D/wide3D}, and \emph{wm2D/wm3D} this is a hydrostatic equilibrium boundary condition on the density that maintains hydrostatic equilibrium by assuming constant internal energy and constant radial acceleration due to gravity in the boundary cells \citep{hsegrimm2015realistic}.    For simulations \emph{dcs2D/dcs3D} and \emph{cc2D/cc3D} the outer radial boundary has a constant radial derivative imposed on the density.  For simulations \emph{bg2D/bg3D}, \emph{wide2d/wide3d}, \emph{wm2D/wm3D}, and \emph{dcs2D/dcs3D}, we impose periodicity on all physical quantities at the boundaries in $\theta$ and $\phi$. For simulation \emph{cc2D} the angular direction is treated with reflective boundary conditions; for simulation \emph{cc3D} both angular directions are treated with periodic boundary conditions.

\section{Differences between 2D and 3D stellar simulations \label{sec2d3dbench}}

\begin{figure}
\begin{center}
\resizebox{3.5in}{!}{\includegraphics{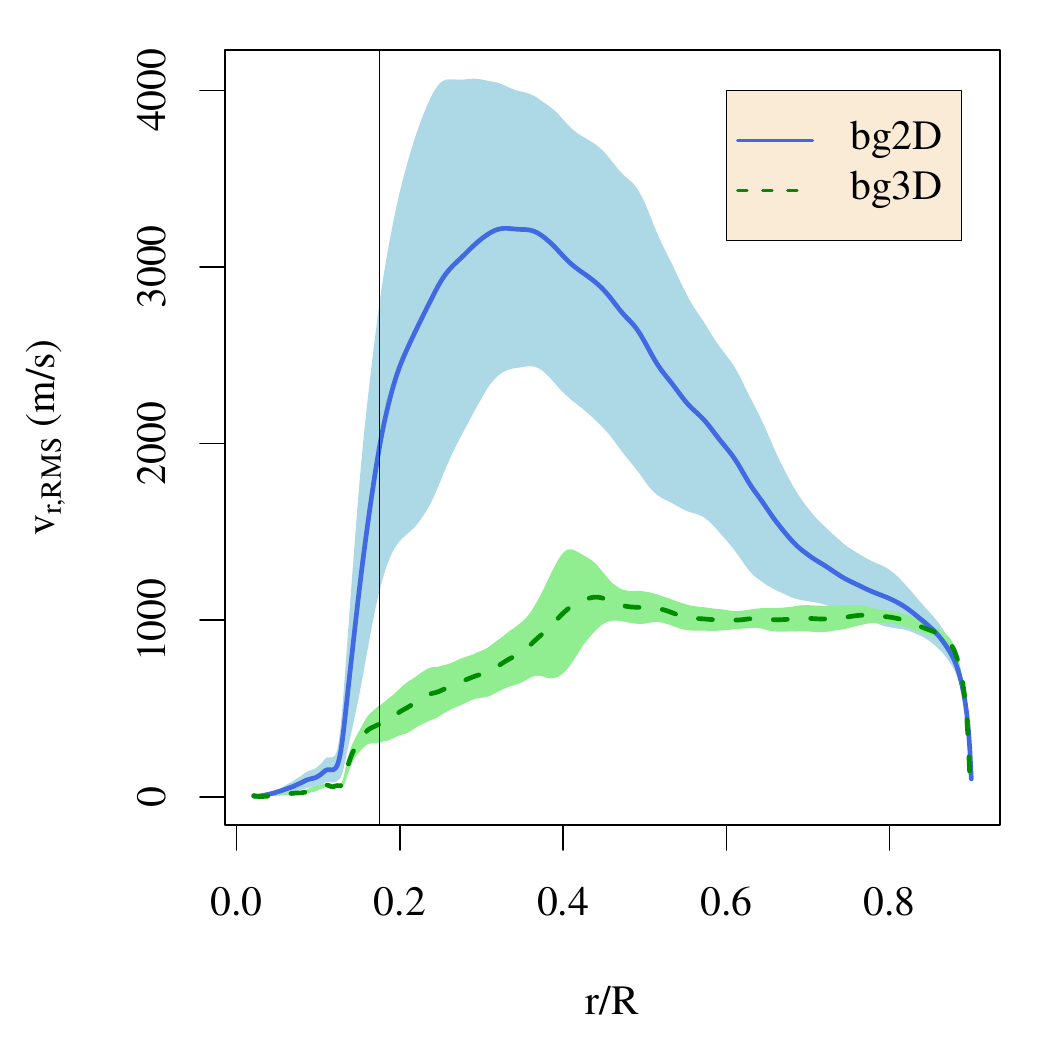}}
\caption{The radial profile of RMS radial velocity $v_{r,\mathsf{RMS}}$ for 2D and 3D simulations of the $3 M_{\odot}$ red giant star.  The lines indicate a time average, taken over the entire simulation time, of the horizontally averaged radial profile. The shaded regions represent one standard deviation above and below the time-averaged line. The radial position of the convective boundary, calculated by the Schwarzschild criterion, is indicated by a vertical black line.  The interior radial coordinate of the star $r$ is normalized by the star's radius $R$. \label{figvel_pairs}}
\end{center}
\end{figure}

\begin{figure*}
\begin{center}
\includegraphics[height=3.5in]{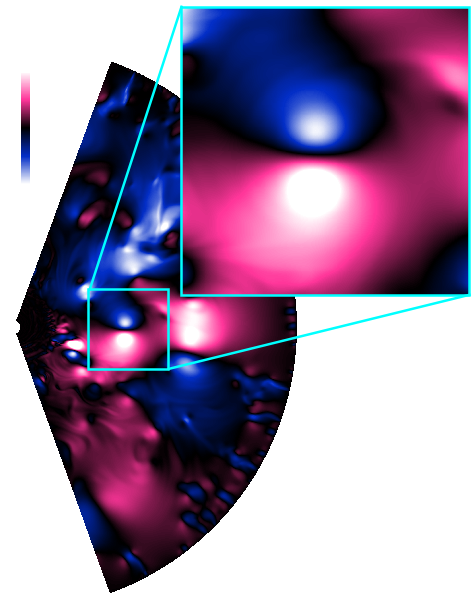}\includegraphics[height=3.5in]{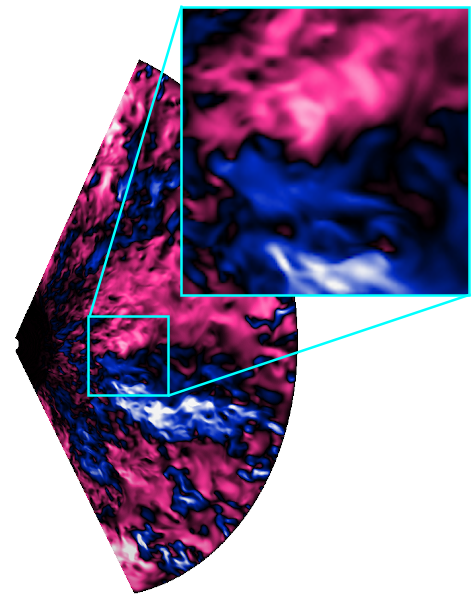}
\caption{{A comparison of radial velocity in the $3 M_{\odot}$ red giant star from (left) 2D simulation \emph{bg2D}, and (right) 3D simulation \emph{bg3D}.  The visualization is zoomed in on a small region inside the convection zone to emphasize the differences in the shape of convection between 2D and 3D.  Outward flows are indicated in pink, while inward flows are in blue; the zero point in velocity is black. The maximum and minimum values of the color scale are defined by a radial velocity magnitude near the maximum for each simulation: \emph{bg2D} ($\pm 6.5$ km/s), \emph{bg3D} ($\pm 2.9$ km/s).} \label{vizvel_pairs}}
\end{center}
\end{figure*}

For each of the pairs of simulations that we study, the root-mean-square (RMS) radial velocity profile in the 2D simulation is larger than in the 3D simulation (see for example Fig.~\ref{figvel_pairs}).  This is a common result for stellar convection also noted in \citet{muthsam1995numerical} and \citet{pratt2020comparison}; the extent to which the 2D velocities are larger appears to be dependent on the stellar model examined.  In addition, the radial velocities in 2D and 3D simulations look different; 3D simulations appear to be ``rougher,'' i.e. more small-scale structure is present.  This is particularly visible at points where inflows and outflows interact (see Fig.~\ref{vizvel_pairs}).  

\begin{figure}
\begin{center}
\resizebox{3.5in}{!}{\includegraphics{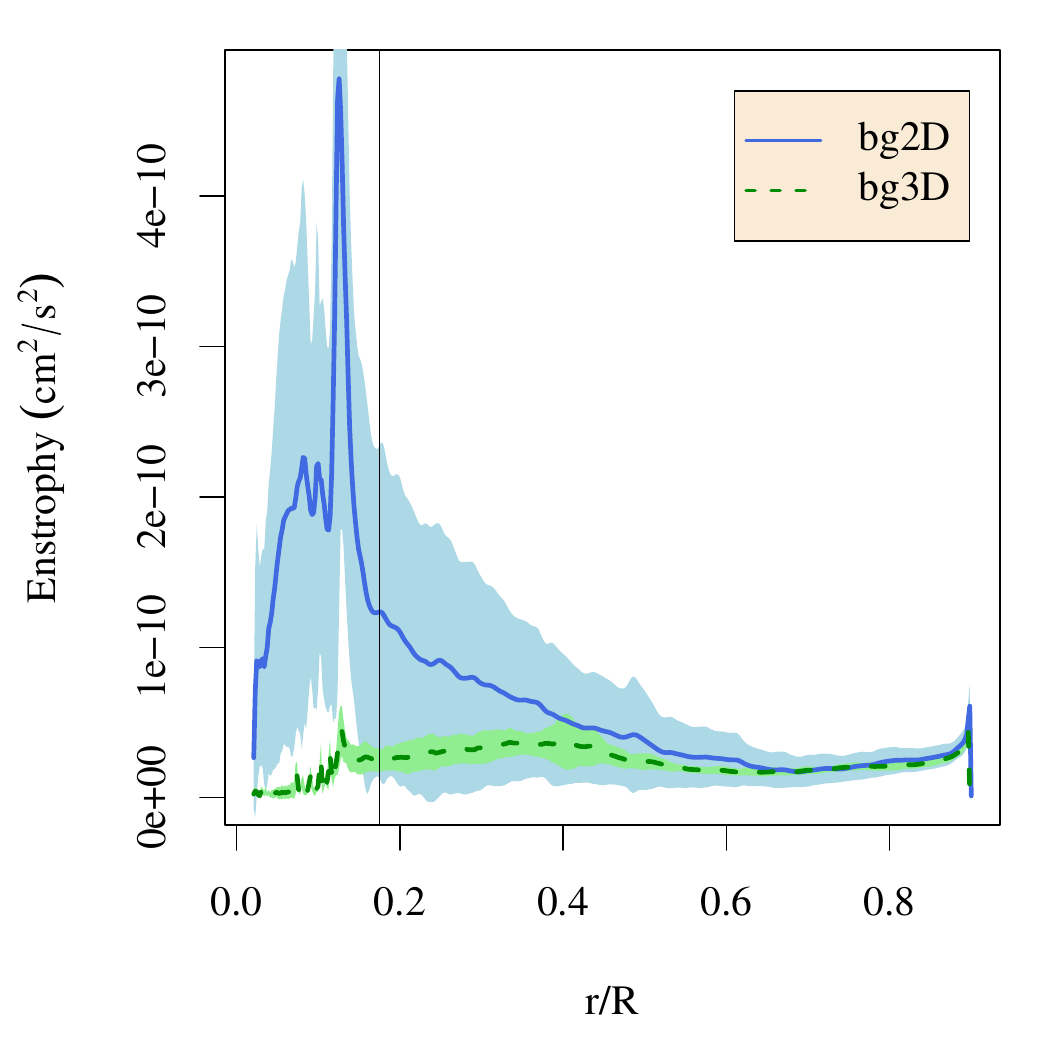}}
\caption{Radial profiles of the local $\phi$ enstrophy for 2D and 3D simulations of the $3 M_{\odot}$ red giant star. The lines indicate a time average, taken over the entire simulation time, of the horizontally averaged radial profile. The shaded region represents one standard deviation above and below the time-averaged line. The radial position of the convective boundary {determined by the Schwarzschild criterion} is indicated by a vertical black line.
\label{figvort_pairs}
}
\end{center}
\end{figure}

The differences between 2D and 3D simulations also reach beyond velocity amplitudes into the structure of the flow.   We examine the radial profile of the local enstrophy in the $\phi$-direction, defined as the square of the $\phi$ component of the vorticity $\omega_{\phi} = \nabla \times \vec{v}|_{\phi}$.  The local $\phi$ enstrophy is larger in 2D than in 3D, with the largest differences occurring at, or near, the convective boundary (see Fig.~\ref{figvort_pairs}).   These plots reveal a different shape and structuring in the flow that occurs near convective boundaries based on the dimensionality.

We calculate the overshooting depth $\ell_{\mathsf{ov}}$ by fitting the distribution of overshooting plumes calculated using the vertical kinetic energy flux with a generalized extreme value distribution, as described in \citet{pratt2017extreme}.   We adopt the location parameter from this fit as the overshooting depth $\ell_{\mathsf{ov}}$. The values of $\ell_{\mathsf{ov}}$ are provided in Table~\ref{table_results} for all of our simulations.  For the simulation pairs \emph{wm2d/wm3D}, \emph{bg2D/bg3D}, and \emph{cc2D/cc3D} these numbers are extremely close.  For the simulation pair \emph{dcs2D/dcs3D}, the 2D simulation has somewhat deeper overshooting, while for the simulation pair \emph{wide2D/wide3D}, the 3D simulation has somewhat deeper overshooting.  Given the limited amount of data for the 3D simulations, it is not clear whether these differences between the 2D and 3D simulations are statistically significant.

\section{Filling factor calculations \label{secresultsid}}

\subsection{Definition of the filling factor based on volume percentage}

We define a volume-percentage filling factor to be the fractional volume occupied by either the inflows $\sigma_{\mathsf{vp,in}}$ or the outflows $\sigma_{\mathsf{vp,out}}$:
\begin{eqnarray}\label{vpfffirst}
\sigma_{\mathsf{vp,in}} &=& \frac{ V^{\mathsf{in}}(r,t)}
{ V^{\mathsf{in}}(r,t) + V^{\mathsf{out}}(r,t) }
\\
\sigma_{\mathsf{vp,out}} &=& \frac{ V^{\mathsf{out}}(r,t)}
{ V^{\mathsf{in}}(r,t) + V^{\mathsf{out}}(r,t) }
\end{eqnarray}
Here $V^{\mathsf{in}}$ indicates the total volume of grid cells at a given radius that has an inward velocity, while $V^{\mathsf{out}}$ indicates the total volume of grid cells that have an outward velocity.  The natural consequence is that the sum of the filling factors of inflows and outflows must be one $\sigma_{\mathsf{vp,in}}+ \sigma_{\mathsf{vp,out}}=1$. 
Because of this relation, we will generally use the notation $\sigma_{\mathsf{vp}}$ to indicate the filling factor for the plumes moving toward the convective boundary, dropping the ``in'' and ``out'' labels.   Conceptually, the volume-percentage filling factor is equating the situation where there are many small plumes with an equivalent single large inflow and single large outflow. 
Many works \citep{schmitt1984overshoot,brummell2002penetration,andrassythesis,kapyla2023convective} have used this kind of definition for a filling factor.  Some have chosen to evaluate the area at the cell surfaces rather than the volume of the cell to calculate this filling fact.  These two alternatives converge toward the same value as the cell size is decreased;
using the volume to calculate this filling factor is convenient because \music~is a finite-volume code.   
In the work of \citet{zahn1991convective}, the implication is that the filling factor is a single number, independent of radius.  However other authors, including \citet{schmitt1984overshoot,cai2020upward,canuto1998stellar,kapyla2023convective} formulate a filling factor that is a function of the radial depth in the star; doing so allows us to examine ideas about non-local convection throughout the stellar radius.  
Based on ideas acquired from early simulations of solar convection, \citet{canuto1998stellar} expect a highly asymmetric convection pattern with fast, concentrated inflows and slow, broad outflows. Based on observational data at the solar surface and the idea of a stratified star, they thus assert that the filling factor for inflows is always less than half: $\sigma_{\mathsf{vp,in}} < 1/2$. 

\begin{figure}
\begin{center}
\resizebox{3.5in}{!}{\includegraphics{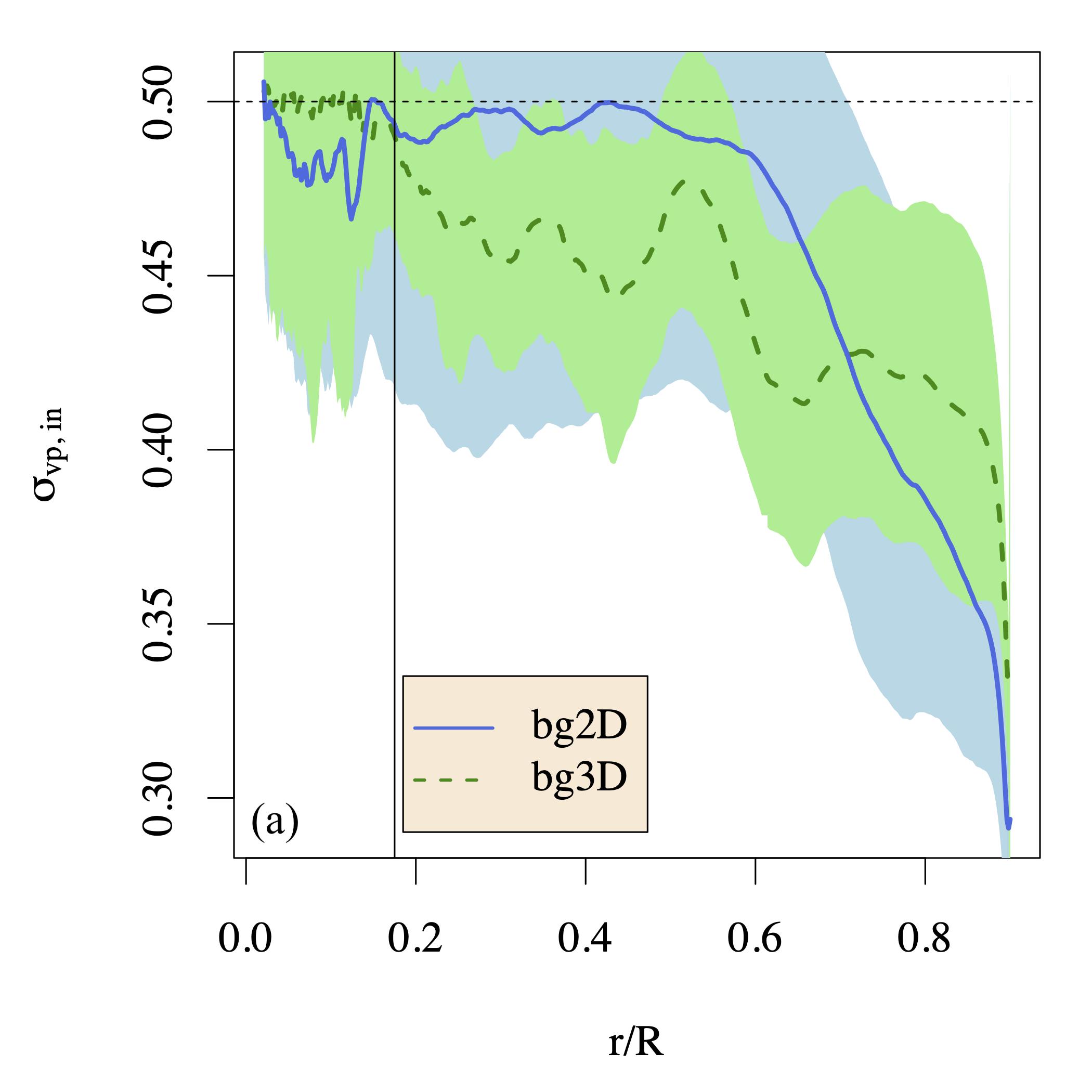}}
\resizebox{3.5in}{!}{\includegraphics{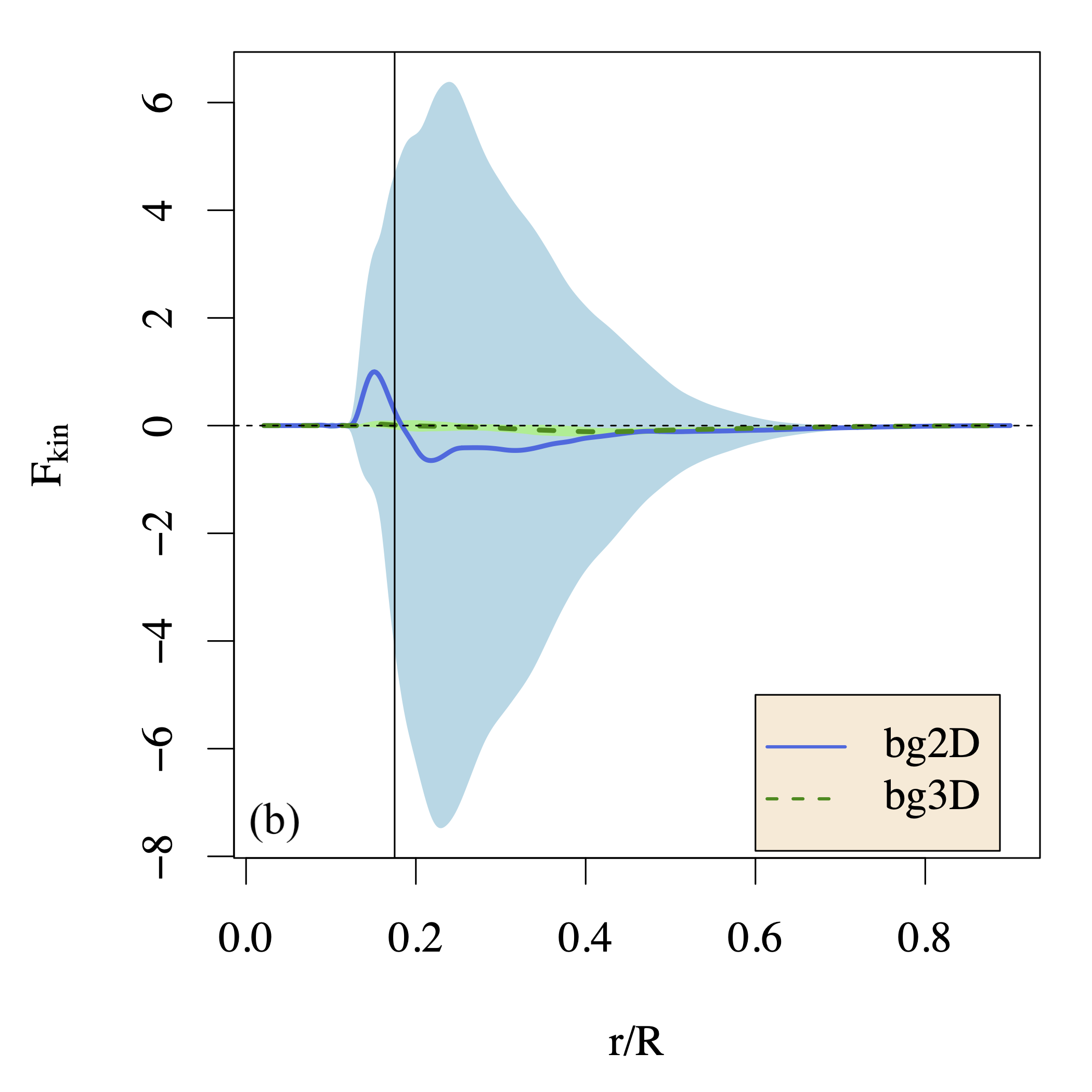}}
\caption{(a) Radial profile of the volume-percentage filling factor of the inward moving plumes $\sigma_{\mathsf{vp,in}}$ vs the star's internal radius, in units of the total stellar radius $R$ for simulations \emph{bg2D/bg3D}.  (b) Radial profile of the normalized kinetic energy flux, $F_{\mathsf{k}}$ for simulations \emph{bg2D/bg3D}.   The solid and dashed lines indicate time-averaged radial profiles. Shaded areas indicate one standard deviation above and below these averaged lines.  A thin vertical line indicates the convective boundary determined by the Schwarzschild criterion.
\label{sigmargb}}
\end{center}
\end{figure}

A calculation of $\sigma_{\mathsf{vp}}$ in our simulation pair \emph{bg2D/bg3D} is shown in Fig.~\ref{sigmargb}(a).   Both of these simulations evidence a $\sigma_{\mathsf{vp}}$ close to a third near the surface of the star, in rough agreement with observations of solar surface convection \citep[e.g.][]{nordlund2009solar}.  The similarity between our simulations and observations of the solar surface is striking, given that the resolution of near-surface dynamics is challenging for simulations of the stellar interior.  As we examine $\sigma_{\mathsf{vp}}$ deeper in the convective envelope, we find that it grows to approximately one-half, indicating that the convection becomes highly symmetric at the point that plumes are overshooting the bottom of the convection zone.  This result is interesting when considered in conjunction with the kinetic energy flux (see Fig.~\ref{sigmargb}(b)).   The kinetic energy flux in all of our simulations is negative in the upper and middle parts of the convection zone. It becomes positive near the convective boundary, a result not seen in early works \citep[e.g.][]{hurlburt1984two}.   These combined results indicate that flows in the convection zone are more complex than the simple picture of thin, fast-moving inward plumes.  In the highly stratified deep interior, inward moving plumes can be both faster and wider than plumes in the near-surface layers.

\begin{figure}
\begin{center}
\resizebox{3.5in}{!}{\includegraphics{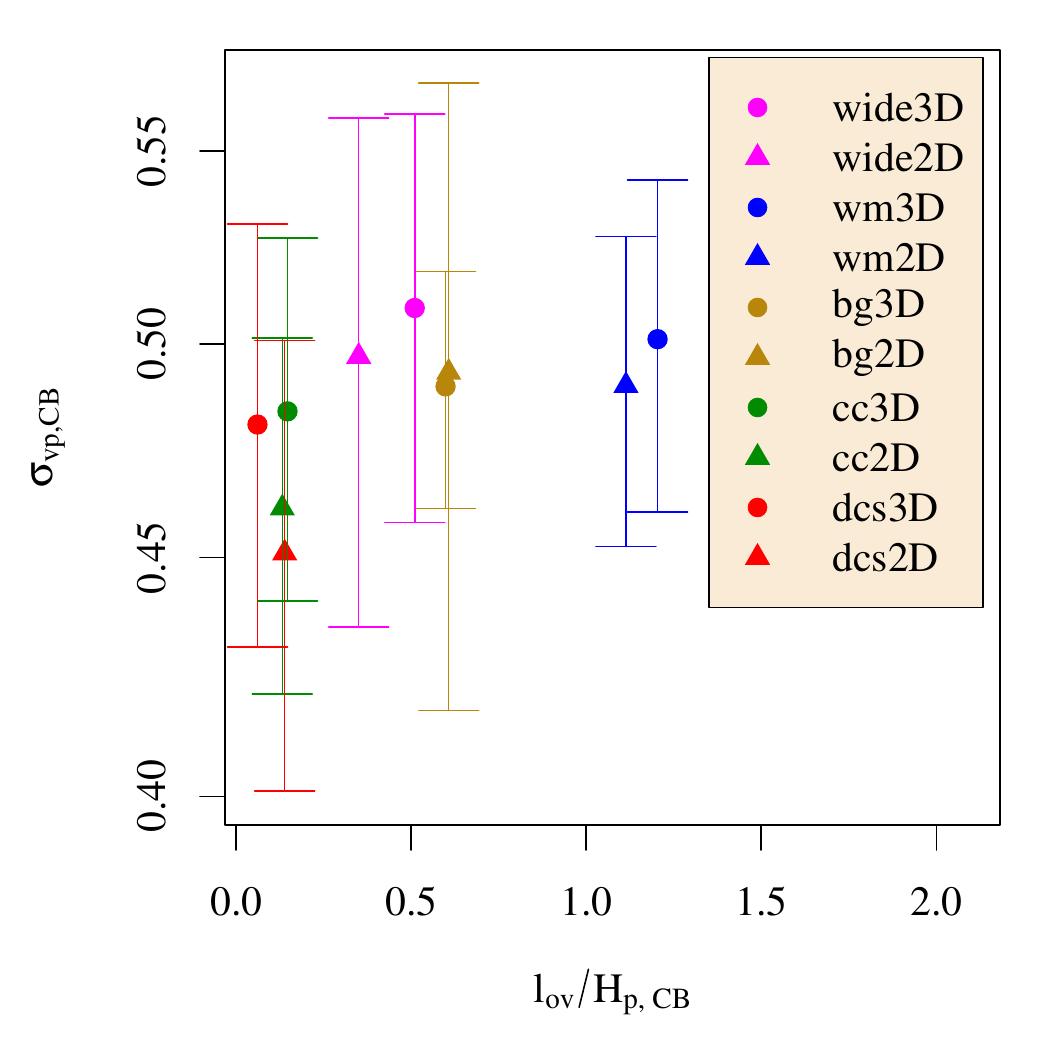}}
\caption{The time-averaged volume-percentage filling factor evaluated at the convective boundary {determined by the Schwarzschild criterion}, $\sigma_{\mathsf{vp,CB}}$, vs the overshooting depth in units of the pressure scale height at the convective boundary  {$\ell_{\mathsf{ov}}/H_{p,CB}$} for all simulations described in Table~\ref{simsuma}.   Error bars indicate one standard deviation around the time-averaged value of $\sigma_{\mathsf{vp,CB}}$.  Error bars obtained in the calculation of the overshooting depth $\ell_{\mathsf{ov}}$ are smaller than the size of the data points.
\label{figsigmavslov}}
\end{center}
\end{figure}

We calculate the value of the volume-percentage filling factor for plumes moving toward the convective boundary in each simulation.  We then extract the value at the convective boundary, $\sigma_{\mathsf{vp, CB}}$, for each of our simulations; these can be found in Table \ref{table_results}.  Across all simulations, the mean value of $\sigma_{\mathsf{vp, CB}}$ is 0.499 and the median value is 0.50.  For each simulation, the value of 0.5 is within one standard deviation of the time-averaged value.   We find no clear difference between the values of $\sigma_{\mathsf{vp, CB}}$ calculated from 2D and 3D simulations.  The distributions of these values, calculated at different points in time during steady state convection, strongly overlap; for some simulations, the time-averaged 2D value is slightly larger, and for others, the time-averaged 3D value is slightly larger.  Given the wide range in overshooting depths that we calculate for the four different stars that we examine in this work, we find no clear correlation between the volume-percentage filling factor at the convective boundary and the overshooting length (see Fig.~\ref{figsigmavslov}). 

\begin{table*}
\begin{center}
\caption{Calculated quantities related to the filling factor for convection. The table includes: 
the overshooting depth $\ell_{\mathsf{ov}}$ in units of the pressure scale height at the convective boundary, 
 the volume-percentage filling factor $\sigma_{\mathsf{vp, CB}}$, the mass-flux filling factor $\sigma_{\mathsf{mf, CB}}$, and the incompressible convective-flux filling factor $f_{\mathsf{z, CB}}$.  The subscript CB indicates that the quantity is evaluated at the convective boundary (CB), as defined by the Schwarzschild criterion. Several additional quantities that are evaluated above or below {this} convective boundary are also shown.  This includes the Anders penetration parameter $P_{\mathsf{A}}$ and the plume interaction parameter $\sigma_{\mathsf{int}}$.  It also includes for convective envelopes  (convective cores) the average number of inflows (outflows) in the overshooting layer $N_{\mathsf{OL}}$ and in the convection zone $N_{\mathsf{CZ}}$,  and the average width of inflows {(outflows)} in the overshooting layer $W_{\mathsf{OL}}$ and in the convection zone $W_{\mathsf{CZ}}$.  The average widths are displayed in units of the percentage of the simulation volume at the given radius.
 \label{table_results}
 }
\begin{tabular}{lcccccccccccccccccccccccccccc}
                       & $\ell_{\mathsf{ov,k}}/H_{\mathsf{p,CB}}$ & $\sigma_{\mathsf{vp, CB}}$  & $\sigma_{\mathsf{mf, CB}}$ &  $f_{z,\mathsf{CB}}$  & $P_{\mathsf{A}}$  &  $N_{\mathsf{OL}}$ & $N_{\mathsf{CZ}}$ & $W_{\mathsf{OL}}$ &
                       $W_{\mathsf{CZ}}$ & $\sigma_{\mathsf{int}}$
                       
\\ \hline
wide3D           & $0.51$   &  0.51  &    0.49        &  -0.11   &  0.25  & 16.0 &  7.6     &  2.8  & 5.3 & 0.53 
\\ \hline
wide2D               &  $0.35$ & 0.50   &   0.50          &   -0.10   &  0.10  & 14.3  &  6.4  & 3.4    &   7.5 & 0.45
\\ \hline
wm3D             &  $1.20$ &    0.50 &   0.50  &  0.00   &  1.52  &   31.3 & 19.9   &   1.6  & 2.0 & 0.79
\\ \hline
wm2D           &   $1.11$ &   0.49 &     0.51       & -0.02    &  0.27  &  26.0  & 12.9  &  1.9  & 3.8 &0.50 
\\ \hline
bg3D             &  $0.60$  &   0.49 &     0.51       & 0.00     &   0.91  &  14.9  & 10.5  & 3.4 & 4.7 & 0.72
\\ \hline
bg2D              &  $0.61$  &   0.49 &   0.51         &    -0.04    &  0.06  &   12.8 & 3.3  &  3.8  & 16.7 & 0.23
\\ \hline
cc3D               &  $0.15$  &   0.48 &   0.52         &    0.00     &  25.8  &   17.7 & 8.7  &  2.6 & 5.0 & 0.51 
\\ \hline
cc2D           &  $0.13$  &  0.46 &   0.54         &    -0.07    &   0.52 &   21.7 & 5.3  &  2.1  & 8.7 & 0.25  
\\ \hline
dcs3D               & 0.06  &  0.48  &  0.52    & 0.00     &  1.88  & 28.4 & 16.0   & 1.5  & 2.3 & 0.67 
\\ \hline
dcs2D          &  {0.14}  &  0.45  &  0.55          &   -0.02   &  0.44   & 23.2 &  8.1   & 2.1  &  6.0 &0.36 
\\ \hline
\end{tabular}
\end{center}
\end{table*}

\subsection{Definition of the filling factor based on mass flux}

We define filling factors based on the vertical mass flux $F_{\mathsf{mass}} = \rho v_r $ so that
\begin{eqnarray}\label{massff}
\sigma_{\mathsf{mf,in}}  &=& \frac{\left|F_{\mathsf{mass}}^{\mathsf{inflows}}\right|}{ \left|F_{\mathsf{mass}}^{\mathsf{inflows}}\right| + \left|F_{\mathsf{mass}}^{\mathsf{outflows}}\right|}~,
\\
\sigma_{\mathsf{mf,out}}  &=& \frac{\left|F_{\mathsf{mass}}^{\mathsf{outflows}}\right|}{ \left|F_{\mathsf{mass}}^{\mathsf{inflows}}\right| + \left|F_{\mathsf{mass}}^{\mathsf{outflows}}\right|}
~.
\end{eqnarray}
The use of absolute values in the denominator of these ratios is necessary because, due to mass conservation, the direct sum of these fluxes is small at any point in time.  We note that
$\sigma_{\mathsf{mf,in}}+ \sigma_{\mathsf{mf,out}}=1$, and we will generally use the notation $\sigma_{\mathsf{mf}}$ to indicate the filling factor for the plumes moving toward the convective boundary.  This definition of a mass-flux filling factor represents a mass-weighted rather than volume-weighted version of the filling factor.

\begin{figure}
\begin{center}
\resizebox{3.5in}{!}{\includegraphics{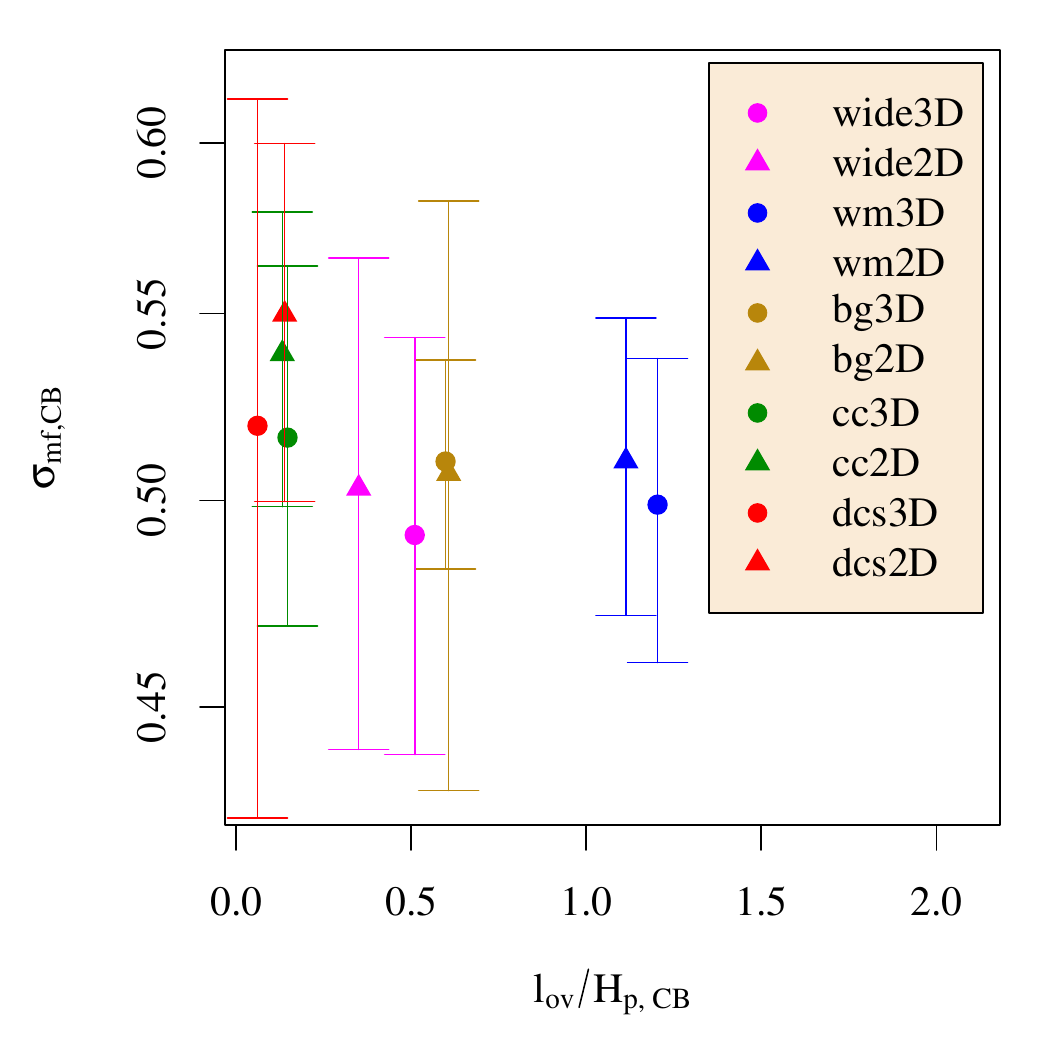}}
\caption{The time-averaged mass-flux filling factor evaluated at the convective boundary {determined by the Schwarzschild criterion} $\sigma_{\mathsf{mf,CB}}$ vs the overshooting depth in units of the pressure scale height at the convective boundary {$\ell_{\mathsf{ov}}/H_{p,CB}$} for all simulations described in Table~\ref{simsuma}.  Error bars indicate one standard deviation around the time-averaged value of $\sigma_{\mathsf{mf,CB}}$.  Error bars obtained in the calculation of the overshooting depth $\ell_{\mathsf{ov}}$ are smaller than the size of the data points.
\label{sigmassvslov}}
\end{center}
\end{figure}

Table \ref{table_results} includes values of the mass-flux filling factor at the convective boundary, $\sigma_{\mathsf{mf, CB}}$, for each of our simulations.  Like the volume-percentage filling factor, the mass-flux filling factor is close to one-half {at this point}.  Fig.~\ref{sigmassvslov} demonstrates that, like the volume-percentage filling factor, there is no clear trend between 2D and 3D results, with strongly overlapping distributions from 2D and 3D simulations.  There is also no clear correlation between the mass-flux flux filling factor and the overshooting length.  For the selection of stars that we study in this work, the mass-flux filling factor provides similar information to the volume-percentage filling factor.

\citet{canuto1998stellar} defined a filling factor based purely on the velocity (see eq.~(49a) of that work). For the simulations examined in this work, this velocity-based filling factor produces results similar to the mass-flux filling factor.

\subsection{Definition of the filling factor based on convective flux}

The analytical work of \citet{zahn1991convective} is based on the radial profile of the convective flux
\begin{eqnarray}\label{fcon}
\overline{F}_{\mathsf{conv}}(r,t) = - c_p(r,t) \rho(r,t) \overline{u_r(r,\theta,t)  T_1(r,\theta,t) }~.
\end{eqnarray}
The bar on the right-hand side of this equation indicates an average over the horizontal directions, here indicated by $\theta$.  The temperature fluctuation $T_1$ is the difference between the local temperature field $T$ and the average temperature at a given radius, angle, and time.  We conveniently measure this temperature fluctuation as a deviation from an initial radial profile, because the radial profile of the temperature does not evolve during our simulations.   This formulation of the convective flux neglects any changes in the density and specific heat that could be dependent on an angle.

\begin{figure}
\begin{center}
\resizebox{3.5in}{!}{\includegraphics{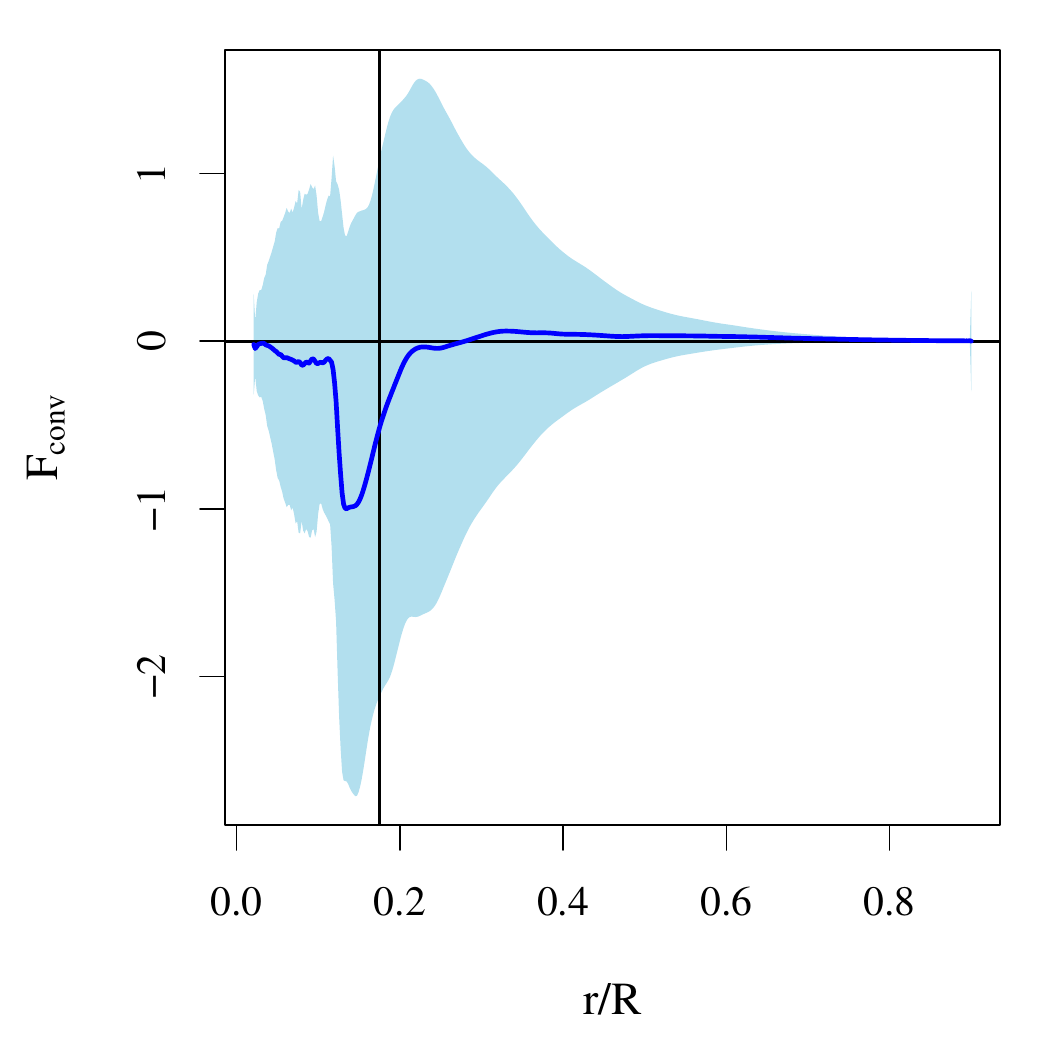}}
\caption{The radial profile of time-averaged convective flux for the $3$ M$_{\odot}$ red giant star \emph{bg2D}, normalized to its maximum magnitude value. The shaded region represents one standard deviation in time, above and below the time-averaged line.  The radial position of the convective boundary {determined by the Schwarzschild criterion} is indicated by a vertical black line.
\label{convfluxradial}}
\end{center}
\end{figure}

Examining the convective flux in eq.~\eqref{fcon}, we find that the difference between this flux in the inflows and outflows is large around the convective boundary, resulting in a characteristic negative peak when the total convective flux is calculated (see Fig.~\ref{convfluxradial}).  This is consistent with earlier works; both \citet{brun2011modeling} and \citet{browning2004simulations}
define the top of the overshooting layer as the point where the time average of the convective flux first changes sign.  They then define the bottom of the overshooting layer as the radius in the radiative zone where the convective flux becomes small. 
 As is clear in the figure, using this definition of an overshooting layer along with our fully compressible simulations would mean defining a sizable portion of the convection zone as part of the overshooting layer.  This shape of the convective flux is not the result of changes in the temperature gradient of the star; the average temperature profile has not evolved during the course of our simulations.  The large extent of the shaded area in the figure also demonstrates that the time-variation in the convective flux is substantial.  These characteristics are universally present for the different kinds of stars we have simulated.

In their derivation of a filling factor based on the convective flux, \citet{zahn1991convective} defines two functions that are ``horizontal structures'' of the fluid $h_1(\theta,t)$ and $h_2(\theta,t)$ such that
\begin{eqnarray}\label{wW}
     u_r(r,\theta,t) &=& u^{\mathsf{in}}_{r,\mathsf{RMS}}(r,t) h_1(\theta,t)~,
\\ \label{tT}
     T_1(r,\theta,t) &=&  T^{\mathsf{in}}_{1,\mathsf{RMS}}(r,t) h_2(\theta,t)~.
\end{eqnarray}
Here $T^{\mathsf{in}}$ and $u^{\mathsf{in}}$ are the temperature fluctuations and radial velocity only in the volume where the fluid is moving radially inward.
We will derive the convective-flux filling factor for the inflows; the corresponding equations for outflows are fully analogous.
The RMS operation takes the average in the horizontal direction, i.e. over $\theta$ if the simulation is 2D, and over both $\theta$ and $\phi$ if the simulation is 3D.   
\citet{zahn1991convective} assumes that $h_1 = h_2 \equiv h$.  To satisfy this assumption requires that the temperature fluctuation and radial velocity are strongly correlated for inflows, or  
\begin{eqnarray}\label{zahnassumption}
     \frac{u_r(r,\theta,t)}{ u^{\mathsf{in}}_{r,\mathsf{RMS}}(r,t)} = \frac{ T_1(r,\theta,t)}{T^{\mathsf{in}}_{1,\mathsf{RMS}}(r,t)}~.
\end{eqnarray}
If an average over a sufficiently long time period is taken, then the numerators on both sides of this equation will be zero.
However, it is unclear how well this assumption is satisfied at any point in time.  As the width of a plume changes with the radius, the ratios in eq.~\eqref{zahnassumption} could change
as well; this could happen if the inflows are less coherent, or if a convective boundary layer alters the flow near the bottom of the convection zone.  The degree to which eq.~\eqref{zahnassumption} is satisfied could also be different for different stars.

\citet{zahn1991convective} then defines the convective-flux filling factor $f_\mathsf{z}$ as a horizontally averaged function from these horizontal structures
\begin{eqnarray}\label{ffdef3}
f_\mathsf{z}(r,t) \equiv \overline{h_1(r,\theta,t) h_2(r,\theta,t)} \equiv \overline{h^2(r,\theta,t)}~.
\end{eqnarray}
Solving eqs. \eqref{wW} and \eqref{tT} for $h_1(\theta,t)$ and $h_2(\theta,t)$, we express eq. \eqref{ffdef3} as
\begin{eqnarray}\label{ffdefcalc}
f_{\mathsf{z}}(r,t) = \frac{\overline{u_r(r,\theta,t)  T_1(r,\theta,t)}}{u^{\mathsf{in}}_{r,\mathsf{RMS}}(r,t) T^{\mathsf{in}}_{1,\mathsf{RMS}}(r,t)} ~.
\end{eqnarray}
This expression clarifies that the filling factor defined by \citet{zahn1991convective} is essentially a horizontal average of the convective flux, normalized by a proxy for the convective flux in the inflows.  The formula in eq.~\eqref{ffdefcalc} is in a convenient form for direct calculations.
Combining this expression for $f_{\mathsf{z}}$ with Zahn's expression for incompressible convective flux in eq. \eqref{fcon}, we represent the convective flux in terms of the convective flux filling factor
\begin{eqnarray}
    \overline{F}_{\mathsf{conv}}(r,t)  &=& - c_p(r,t) \rho(r,t)u^{\mathsf{in}}_{r,\mathsf{RMS}}(r,t) T^{\mathsf{in}}_{1,\mathsf{RMS}}(r,t) f_{\mathsf{z}}(t)~.
\end{eqnarray}
We find that $f_{\mathsf{z}}(t)$ is highly variable in time, and that a long-time average is required to produce a smooth profile; this is unsurprising because this filling factor is related to $F_{\mathsf{conv}}$, which also requires a long-time average to converge.  Our formulation of the convective-flux filling factor is dependent on radius, a departure from the formula written by \citet{zahn1991convective}, where the horizontal structure functions are not radially dependent. 

\begin{figure}[H]
\begin{center}
\resizebox{3.5in}{!}{\includegraphics{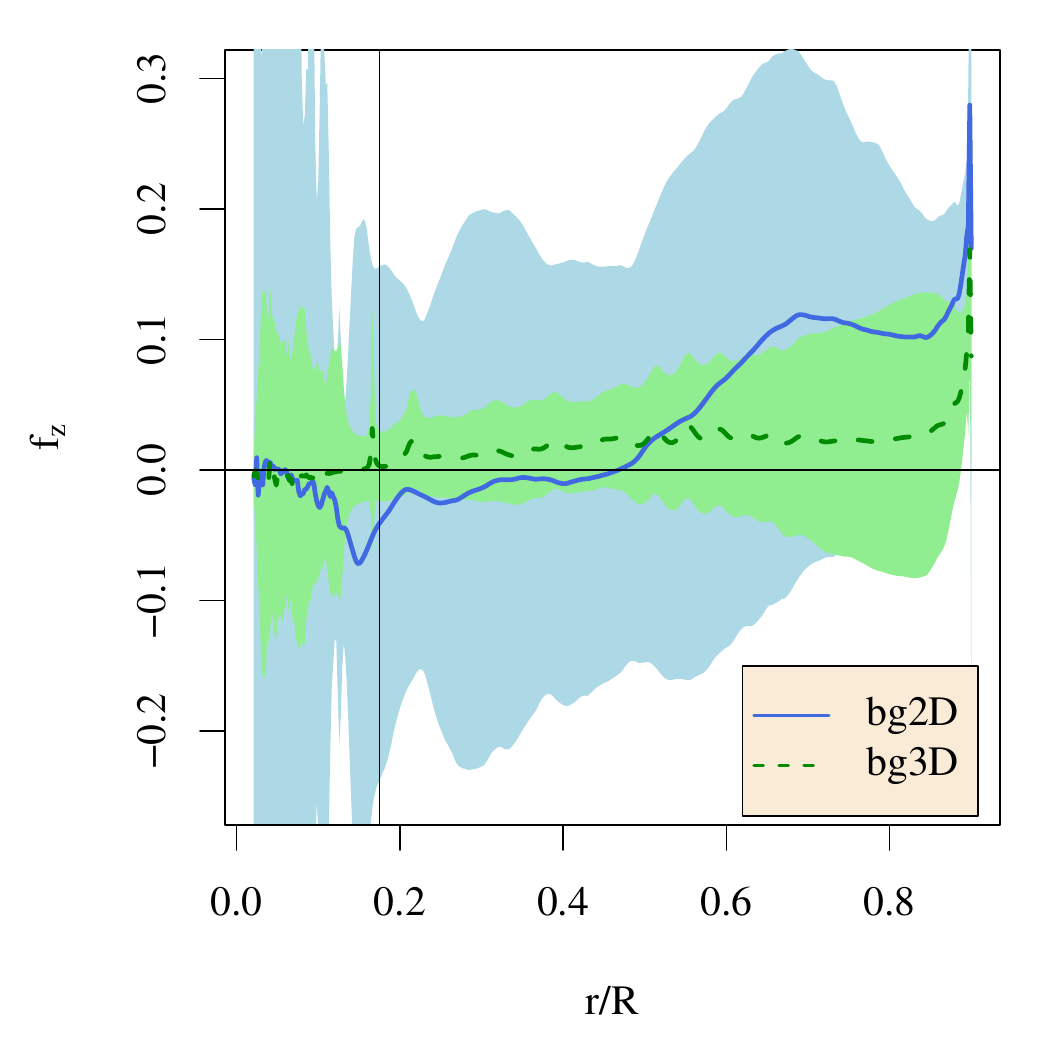}}
\caption{The radial profile of the time-averaged filling factor $f_{\mathsf{z}}$ as defined in eq.~\eqref{ffdefcalc} for 2D and 3D simulations of the $3 M_{\odot}$ red giant star. The lines indicate a time average, taken over the entire simulation time, of the horizontally averaged radial profile. The shaded regions represent one standard deviation above and below the time-averaged line.  The radial position of the convective boundary {determined by the Schwarzschild criterion} is indicated by a vertical black line.
\label{ffradial}}
\end{center}
\end{figure}

A characteristic result for the time-averaged radial profile of the convective-flux filling factor is shown in Fig.~\ref{ffradial}.    Far from the convective boundary, this filling factor is larger and positive.    However, approaching the convective boundary, the convective-flux filling factor becomes small and/or negative.  This appears to be related to the negative peak in the convective flux at the convective boundary.  The implication is that the complex flow patterns, and thus complex fluxes that occur around this boundary contaminate the convective-flux filling factor. A full description of those flows is beyond the scope of the present work, but will be pursued in the future.  For a filling factor, which is conceptualized as a percentage of the flow moving inward, a negative number makes little sense.  Nevertheless, we have also documented the time-averaged value of $f_{\mathsf{z}}$ at the Schwarzschild boundary for convective instability for all of the simulations examined in this work in Table \ref{table_results}.

\subsection{Eliminating compressibility as a source of error for the convective flux filling factor}

\begin{figure}[H]
\begin{center}
\resizebox{3.5in}{!}{\includegraphics{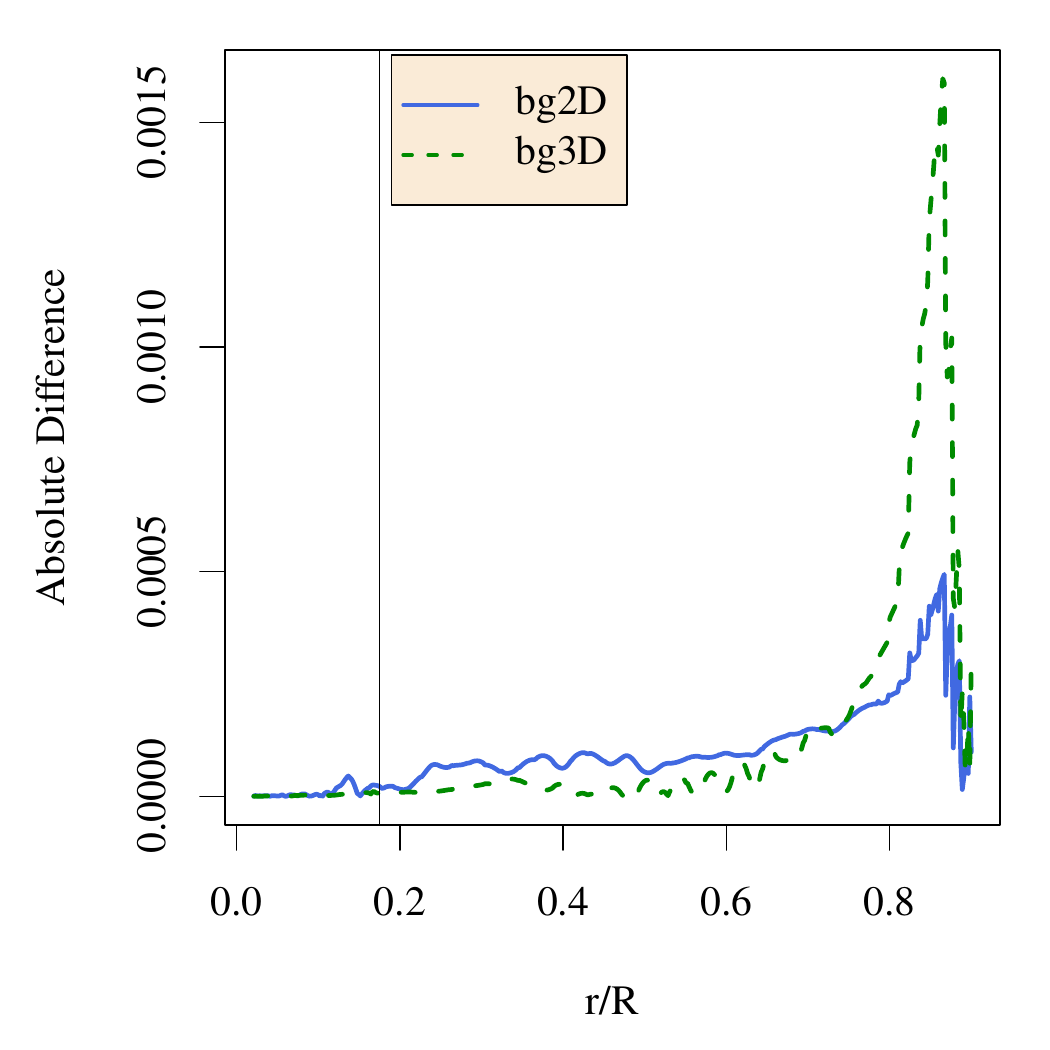}}
\caption{Radial profile of the absolute difference $|f_{\mathsf{z}}-f_{\mathsf{comp}}|$ between the incompressible convective-flux filling factor and the compressible convective-flux filling factor for 2D and 3D simulations of the $3 M_{\odot}$ red giant star. The radial position of the convective boundary {determined by the Schwarzschild criterion} is indicated by a vertical black line.
 \label{ffdiffpercent}}
\end{center}
\end{figure}

One possible source of error in the convective-flux filling factor formulated by \citet{zahn1991convective} is that the assumption of incompressibility could be unphysical around the convective boundary of a star; we therefore expand on Zahn's incompressible definition of a filling factor.  A definition of the horizontally-averaged convective flux that includes compressibility is
\begin{eqnarray}\label{fconalt}
\overline{F}_{\mathsf{conv}}(r,t)  = - \overline{c_p(r,\theta,t) \rho(r,\theta,t) u_r(r,\theta,t)  T_1(r,\theta,t) }~.
\end{eqnarray}
We note that \citet{kapyla2017extended} use a similar definition of this flux where the horizontal average does not include the specific heat.
In this case, we define four horizontal structures such that
\begin{eqnarray}\label{wW1}
     u_r(r,\theta,t) &=& u^{\mathsf{in}}_{r,\mathsf{RMS}}(r,t) h_1(\theta,t)~,
\\ \label{tT1}
     T_1(r,\theta,t) &=&  T^{\mathsf{in}}_{1,\mathsf{RMS}}(r,t) h_2(\theta,t)~,
\\ \label{rhorhop}
     \rho(r,\theta,t) &=& \rho^{\mathsf{in}}_{\mathsf{RMS}}(r,t) h_3(\theta,t)~,
\\ \label{cphor}
     c_p(r,\theta,t) &=&  c^{\mathsf{in}}_{p,\mathsf{RMS}}(r,t) h_4(\theta,t)~.
\end{eqnarray}
As in the incompressible case, the superscript indicates that the RMS only includes contributions where the fluid is moving radially inward, the RMS operation takes the average in the horizontal direction, and we will derive  the filling factor for inward plumes.   We then define a compressible-convective-flux filling factor
\begin{eqnarray}\label{ffdef4}
f_{\mathsf{comp}}(t) &=& \overline{h_1(\theta,t) h_2(\theta,t) h_3(\theta,t) h_4(\theta,t)}~,
\\
 &=& \frac{\overline{c_p(r,\theta,t) \rho(r,\theta,t) u_r(r,\theta,t)  T_1(r,\theta,t)}}{c^{\mathsf{in}}_{p,\mathsf{RMS}}(r,t) \rho^{\mathsf{in}}_{\mathsf{RMS}}(r,t) u^{\mathsf{in}}_{r,\mathsf{RMS}}(r,t) T^{\mathsf{in}}_{1,\mathsf{RMS}}(r,t)} ~.
\end{eqnarray}
We calculate $f_{\mathsf{comp}}$ for all of the simulations studied in this work.  The absolute difference between $f_{\mathsf{z}}$ and $f_{\mathsf{comp}}$ is small near the convective boundary in all of our simulations (see Fig.~ \ref{ffdiffpercent}). The compressible filling factor $f_{\mathsf{comp}}$ is slightly larger near the surface in both 2D and 3D, where compressible effects, e.g. Mach numbers, are expected to be slightly larger; this is typical of all of our simulations. Including the effect of compressibility does not resolve the problematic aspects of the convective-flux filling factor near convective boundaries.

\subsection{The Anders penetration parameter \label{secpp}}

\begin{figure}
\begin{center}
\resizebox{3.5in}{!}{\includegraphics{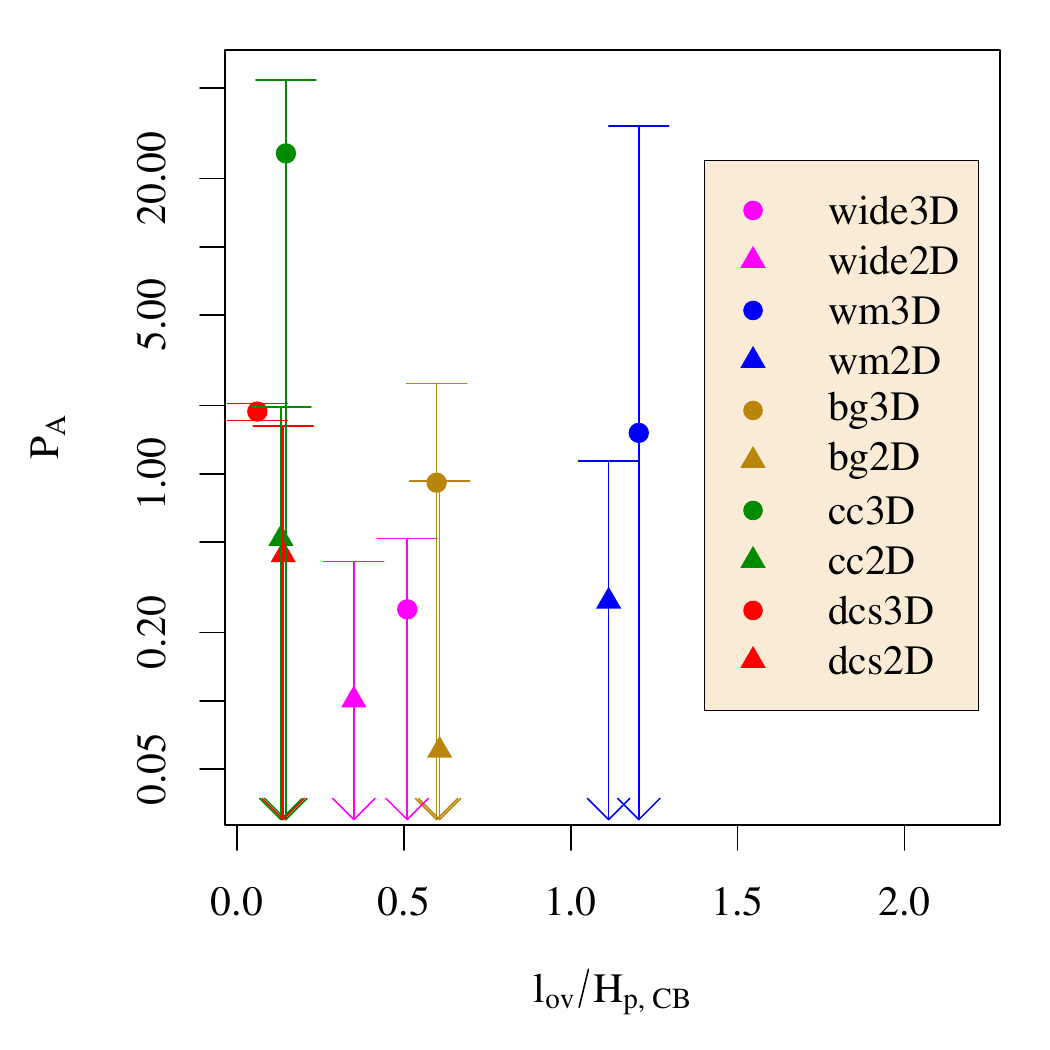}}
\caption{The Anders penetration parameter $P_{\mathsf{A}}$  vs the overshooting depth in units of the pressure scale height at the convective boundary {determined by the Schwarzschild criterion} {$\ell_{\mathsf{ov}}/H_{p,CB}$} for all 2D and 3D simulations described in Table~\ref{simsuma}.  Error bars consider one standard deviation of the convective flux at each of the two radial points that contribute to this number. A logarithmic scale is applied to the vertical axis because this parameter, as formulated, can be larger than one.
\label{lovvspanders}}
\end{center}
\end{figure}

Although it is not strictly a filling factor, we also examine a nondimensional number formulated by \citet{anders2022stellar}, called the ``penetration parameter.''   Similar to \citet{zahn1991convective}, their motivation is to predict convective overshooting and penetration using a ratio of convective fluxes.  It is therefore fitting to examine it here as a comparison to the filling factor in our realistic global simulations of stars.  The Anders penetration parameter $P_{\mathsf{A}}$ is defined as the ratio of the time-averaged convective flux on either side of the convective boundary, namely
\begin{eqnarray}\label{PPdef}
P_{\mathsf{A}} = - \frac{\hat{F}_{\mathsf{conv}}|_{\mathsf{CZ}} }{\hat{F}_{\mathsf{conv}}|_{\mathsf{OL}} } ~.
\end{eqnarray}
In  \citet{anders2022stellar}, ${\hat{F}_{\mathsf{conv}}|_{\mathsf{CZ}}}$ is described as the time-averaged convective flux evaluated ``slightly'' inside the convection zone where both outflows and inflows exist, and $\hat{F}_{\mathsf{conv}}|_{\mathsf{OL}}$ is the time-averaged convective flux located ``slightly'' inside the radiative zone, in a layer where only inflows exist.  In global simulations of stars, this language is not specific enough to determine where the convective flux should be evaluated.  In addition, there is the complication of the complex boundary-layer-like flows that we find inside the convective boundary.  To calculate the Anders penetration parameter across global simulations of different stars, we define ${\hat{F}_{\mathsf{conv}}|_{\mathsf{CZ}}}$ as the maximum time-averaged value of ${F}_{\mathsf{conv}}(r)$ in the convective zone, a point that is unaffected by boundary layer flows.  Similarly, we define $\hat{F}_{\mathsf{conv}}|_{\mathsf{OL}}$ as the minimum time-averaged value of ${F}_{\mathsf{conv}}(r)$, which can be found near the convective boundary, in the overshooting layer.  Values of $P_{\mathsf{A}}$ for each of our simulations, calculated in this way, are shown in Table \ref{table_results}.   The Anders penetration parameter has the useful characteristic of producing larger average values in 2D than in 3D for all of our simulation pairs (see Fig.~\ref{lovvspanders}).   We use the standard deviation in time of ${\hat{F}_{\mathsf{conv}}|_{\mathsf{CZ}}}$ and $\hat{F}_{\mathsf{conv}}|_{\mathsf{OL}}$ to calculate uncertainties for the penetration parameter; the error bars are large, in most cases overlapping data points from both 2D and 3D simulations.  Whether the differences between 2D and 3D simulations that we observe in the Anders penetration parameter can explain differences in the overshooting depth in 2D and 3D, however, remains unclear (see Fig. \ref{lovvspanders}).  A longer time period of data may also be required to successfully evaluate $P_{\mathsf{A}}$. 

Another useful result is that the Anders penetration parameter clearly produces different values for different stellar models.  For the simulation pair \emph{cc2D/cc3D}, which has small overshooting lengths, the minimum of the convective flux in the overshooting layer is also a small value; consequently, the value of $P_{\mathsf{A}}$ becomes large, exceeding 20 for the 3D simulation.  This large value, in addition to the large error bars shown in Fig. \ref{lovvspanders}, is a result of how the ratio is formulated.   Using $1/P_{\mathsf{A}}$ would tend to produce values less than one, as the formulations of filling factors did.

\section{Plume interactions and overshooting \label{secnewstuff3}}

\subsection{The width of inflows \label{newstuff}}

The essence of the filling factor is as a diagnostic for quantitatively measuring how symmetric or asymmetric inflows and outflows are.  The formulation of a filling factor in this way is linked to the early measurement that convective flows observed on the solar surface are structured into thinner regions with intense inflows and broader regions with slower outflows.  We therefore consider this early idea, and we examine directly how plume widths change with radius in our simulations. As a number that sums up the widths of inflows, the filling factor is related to a low-order statistic.  By instead examining the widths directly, we can pursue higher-order statistics that may be different in 2D and 3D convective flows.

\begin{figure}
\begin{center}
\resizebox{3.5in}{!}{\includegraphics{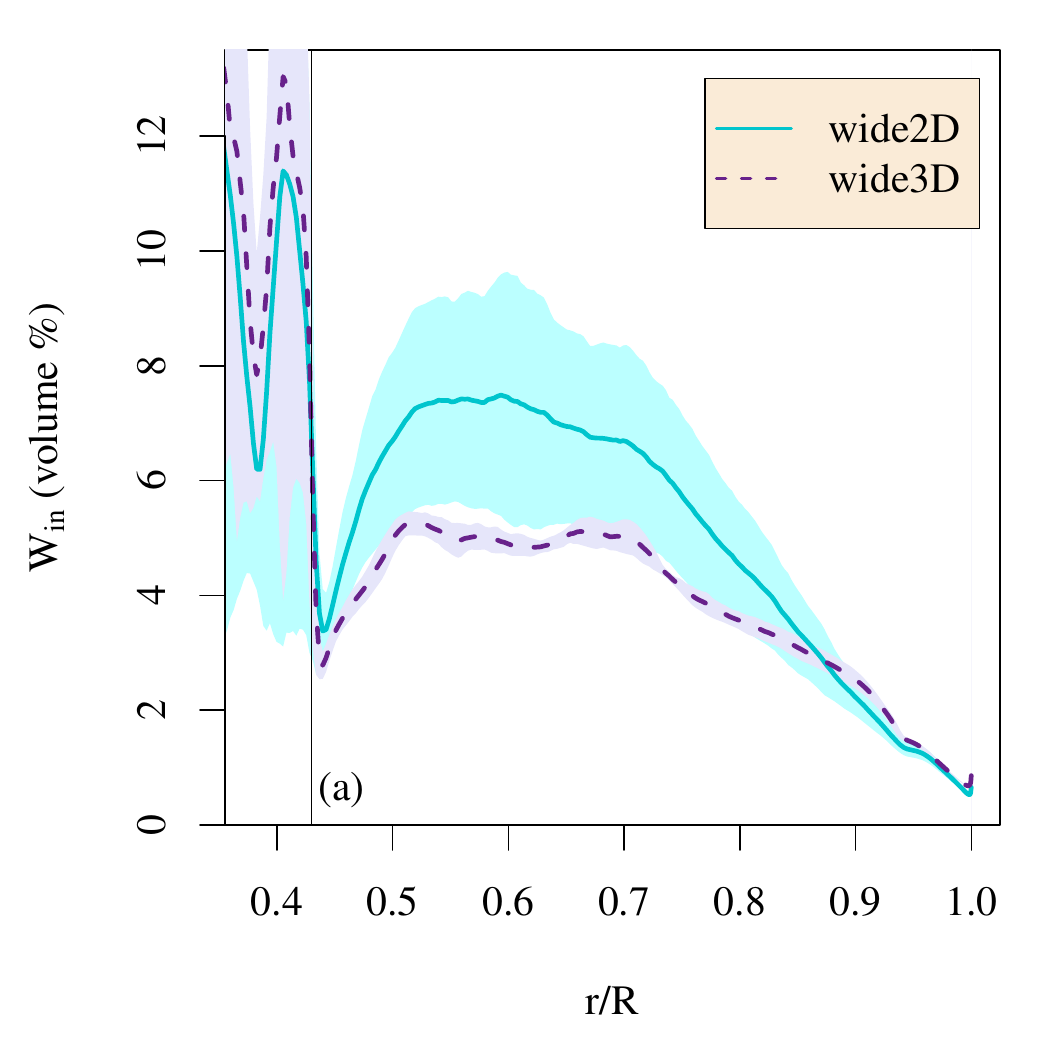}}
\resizebox{3.5in}{!}{\includegraphics{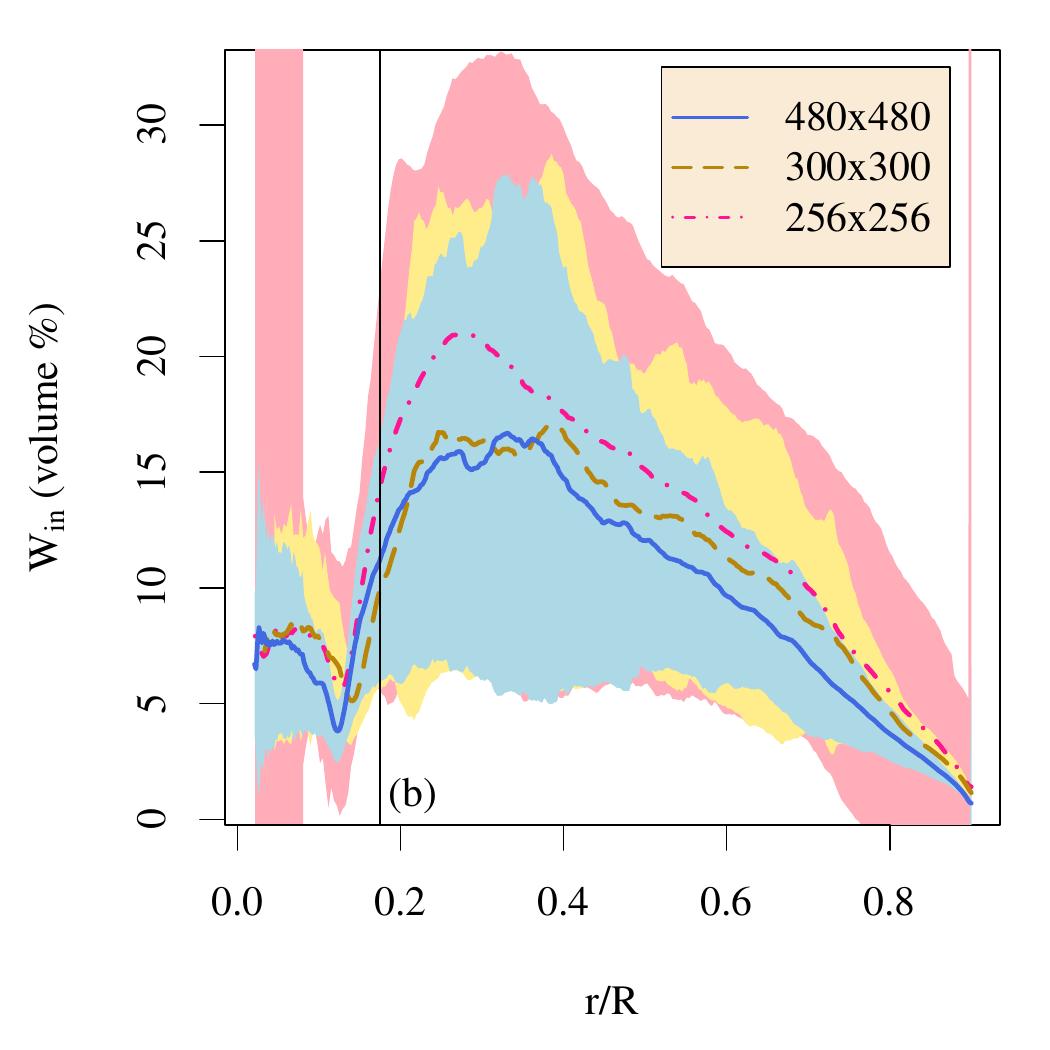}}
\caption{{(a) Widths of inflowing plumes for 2D and 3D simulations of the young sun, \emph{wide2D/3D}.  (b) Widths of inflows for three 2D simulations of the red giant that have different grid sizes as labeled, but are otherwise identical.}  The shaded region represents one standard deviation above and below the time-averaged line.  The radial position of the convective boundary{determined by the Schwarzschild criterion} is indicated by a vertical black line.
\label{winflow_plot}}
\end{center}
\end{figure}

As with the volume-percentage filling factor, we define a single inflow as a continuous set of cells in the $\theta$ direction, at a given radius, that all have a negative radial velocity; 
similarly, we define an outflow based on positive radial velocity. For our 3D simulations, we perform the same calculation for each angle $\phi$ in our grid. 
This allows for the 2D and 3D calculations to be directly compared; otherwise to define a two-dimensional perimeter for a convective plume would require a more involved calculation \citep[e.g.][]{haller2015lagrangian,balasuriya2018generalized,rempel2023lagrangian}.
 For each of our simulations, a characteristic profile is produced for the average widths of plumes as a function of radius (see Fig.~\ref{winflow_plot}).  Two important features emerge in each profile: a maximum average width occurs approximately in the middle of the convection zone, which we call $W_{\mathsf{CZ}}$, and a minimum average width occurs near the bottom of the overshooting layer, which we call $W_{\mathsf{OL}}$.  In our 2D red giant simulation, we find that the difference in the width of plumes between these two points is more than $10\%$.  {Fig.~\ref{winflow_plot} also demonstrates that this characteristic profile is present for simulations with a sufficient resolution; as the grid spacing is decreased, the average width is naturally smaller.}   We find that there is a wider distribution of average widths in the convection zones in each of the 2D simulations that we study than in their 3D counterparts, which may be due to the larger amount of data generated in 2D.

\begin{figure}
\begin{center}
\resizebox{3.5in}{!}{\includegraphics{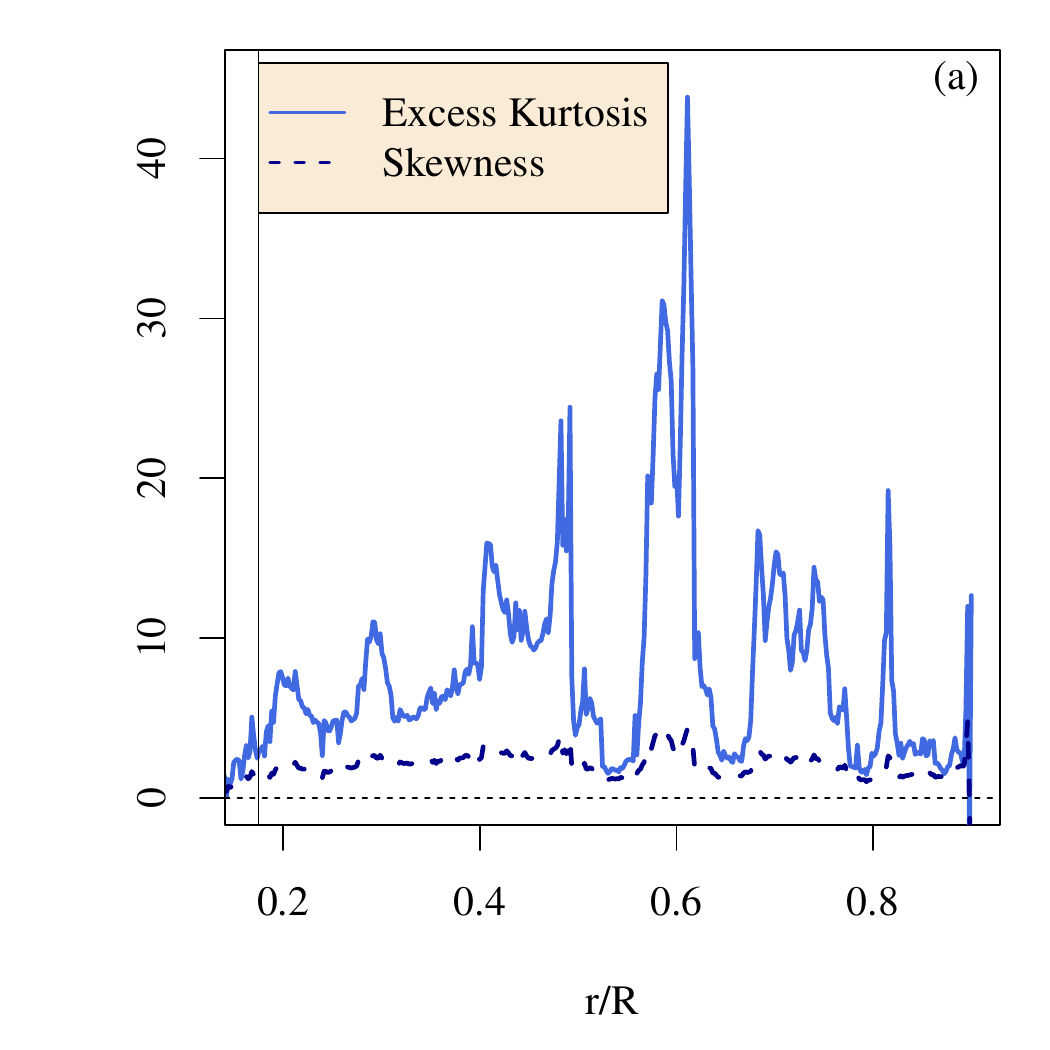}}
\resizebox{3.5in}{!}{\includegraphics{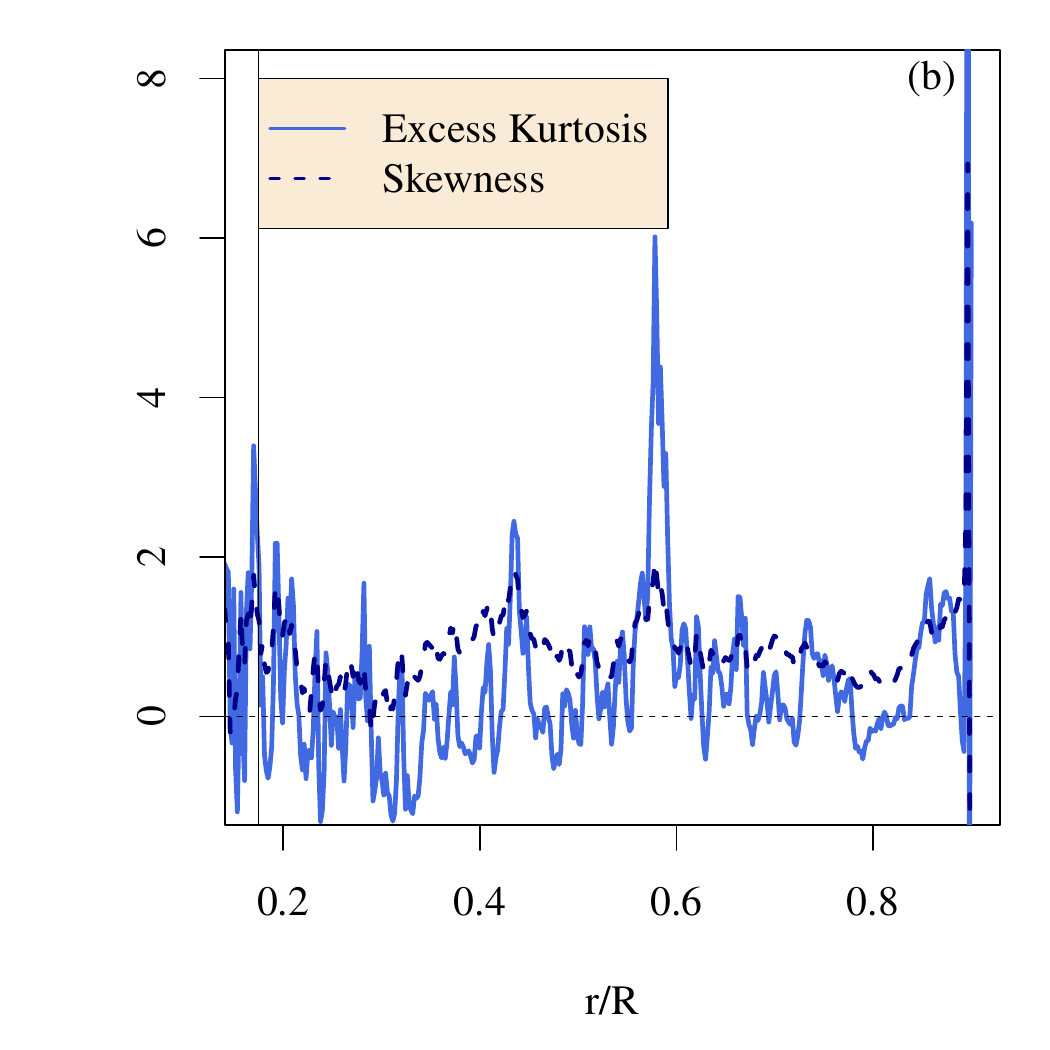}}
\caption{Time-averaged profiles of skewness and excess kurtosis of the width of inflows for the red giant simulations (a) in \emph{bg2D}, and (b) \emph{bg3D}.  The radial position of the convective boundary {determined by the Schwarzschild criterion} is indicated by a vertical black line.
\label{skewinteraction2}}
\end{center}
\end{figure}

\begin{figure}
\begin{center}
\resizebox{3.5in}{!}{\includegraphics{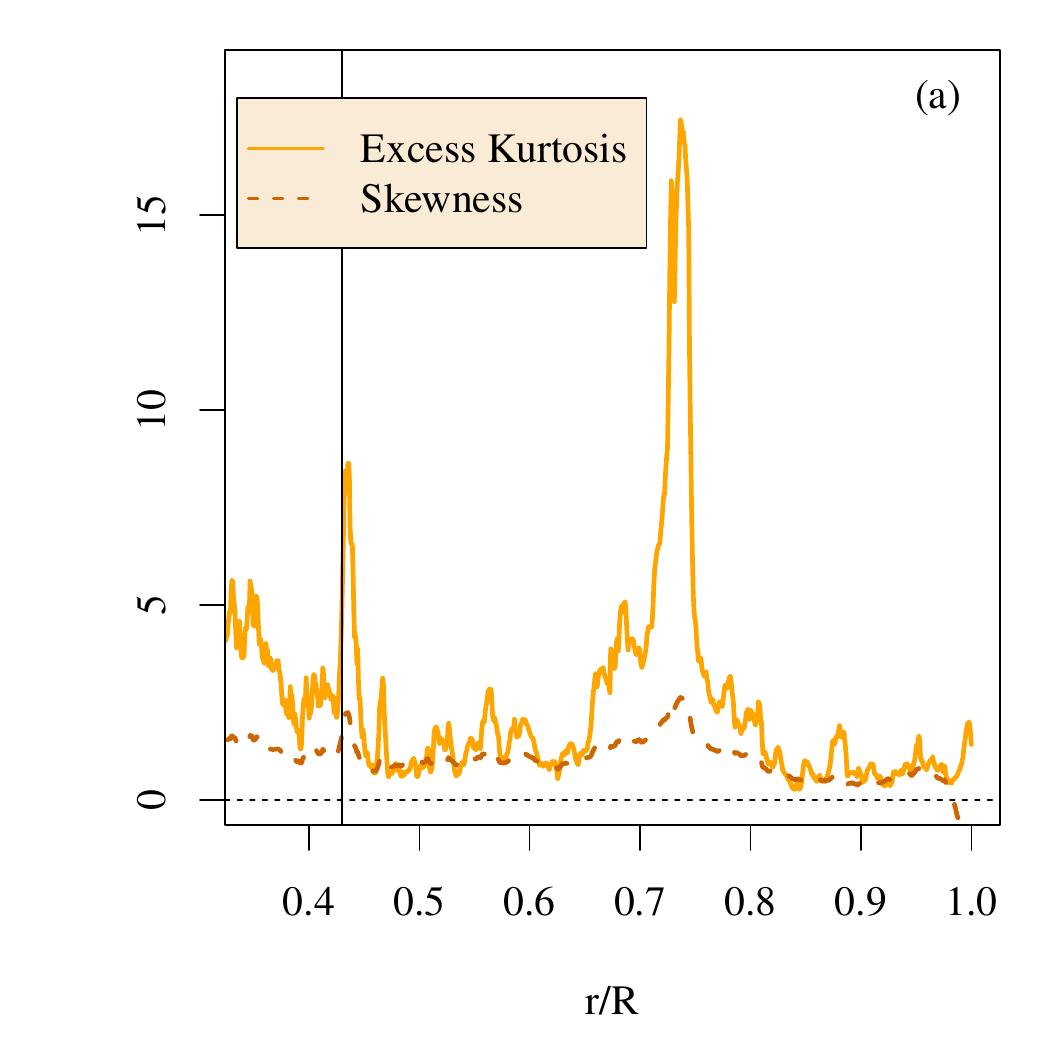}}
\resizebox{3.5in}{!}{\includegraphics{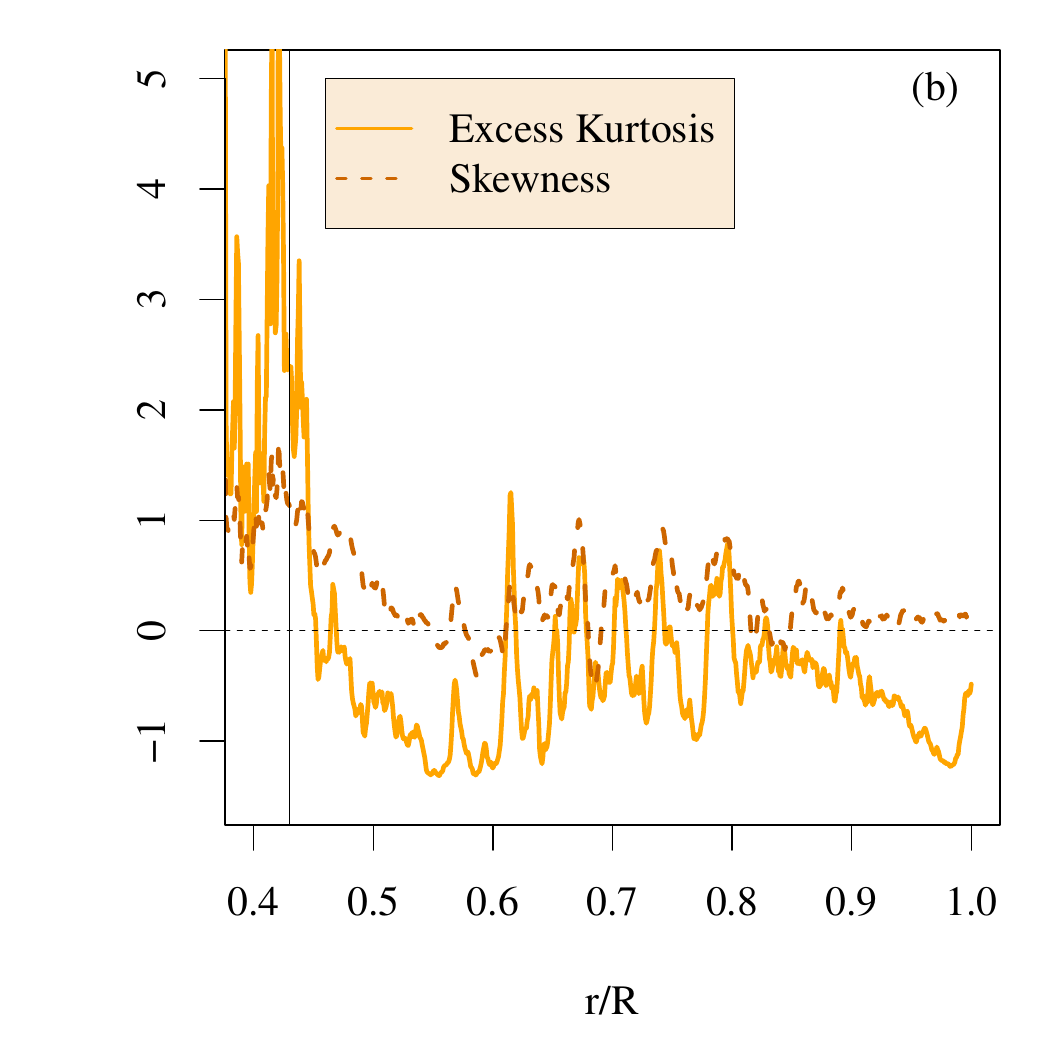}}
\caption{Time-averaged profiles of skewness and excess kurtosis of the width of inflows for the young sun simulations (a) in \emph{wm2D}, and (b) \emph{wm3D}.  The radial position of the convective boundary {determined by the Schwarzschild criterion} is indicated by a vertical black line.
\label{skewinteraction1}}
\end{center}
\end{figure}

The distribution of plume widths in the convection zone is a fundamental diagnostic pointing toward the multi-scale nature of stellar convection.  In comparison, the controlled situation of Rayleigh-B\'enard convection produces convection rolls that have roughly the same size, dictated by the size of the experiment.  In some thin stellar convection zones, this can also be the case; however, in a large convection zone that is defined by a significant density stratification, inflows can have a range of sizes at any given radius. The stellar simulations we examine in this work have convection zones that represent a wide range of stratifications.  Our $20$ M$_{\odot}$ convective core simulation has a density ratio of $\sim 2$ from the top to the bottom of the convection zone, the current sun simulations have a density ratio of $\sim 55$, the red giant simulations include a density ratio of $\sim 200$, and the density ratio in the young sun simulations is greater than $10^{5}$.  Particularly for the young sun, the convection becomes a truly multi-scale flow, with plumes and convection rolls of different sizes frequently interacting with each other.  This is evidenced by the examination of higher-order statistics for plume widths.  The time-averaged radial profile of the skewness of inflow width is generally positive throughout the convection zone in our simulations (see for example Figs.~\ref{skewinteraction2} and \ref{skewinteraction1}), only dipping briefly into the negative for the 3D simulations, where there is a smaller time-series of data available to average.  A normal distribution is defined by a skewness of zero; a positive skewness indicates that the distribution of inflowing plume widths is not symmetrical about the mean value, such that inflowing plumes larger than the average are more prevalent than smaller ones.

These figures also include the excess kurtosis, which is zero for a normal distribution.  In comparison to a Gaussian distribution, a distribution with positive excess kurtosis indicates a greater prevalence of events in the wings of the distribution.  Therefore, the excess kurtosis could give us an indication of the importance of different scales in a convection zone that involves many length scales. In the 2D red giant simulation \emph{bg2D} shown in  Fig.~\ref{skewinteraction2}(a), the excess kurtosis is positive throughout. However, the 3D simulation \emph{bg3D} shown in Fig.~\ref{skewinteraction2}(b) has an excess kurtosis that is both positive and negative. We find similar results for the young sun pair \emph{wm2D/wm3D}; in Fig.~\ref{skewinteraction1}(b), the 3D simulation has negative excess kurtosis. This reinforces the idea that examination of the plume width can expose differences between 2D and 3D simulations.

\subsection{Plume numbers \label{newstuff2}}

The width of inflows can be directly related to their number. The number of inflows does not capture information on asymmetry; however, the number of plumes decreases when inflowing plumes are wider and increases when inflowing plumes are thinner. Some theoretical predictions of overshooting have used the number of plumes \citep{rieutord1995turbulent,pinccon2016generation}.
In addition to this relationship to the plume width, the number of plumes can be related to two possible paradigms for convection in stellar interiors \citep{kapyla2017extended,brandenburg2016stellar,spruit1996convection}.  The first has been described as a ``tree-like'' structure, where the number of inflows is dependent on depth.  The second has been described as a ``forest-like'' structure where the number of inflows is depth-independent. 
To connect these ideas to overshooting, and the differences between 2D and 3D convection, we examine numbers of plumes in our simulations.

\begin{figure}
\begin{center}
\resizebox{3.5in}{!}{\includegraphics{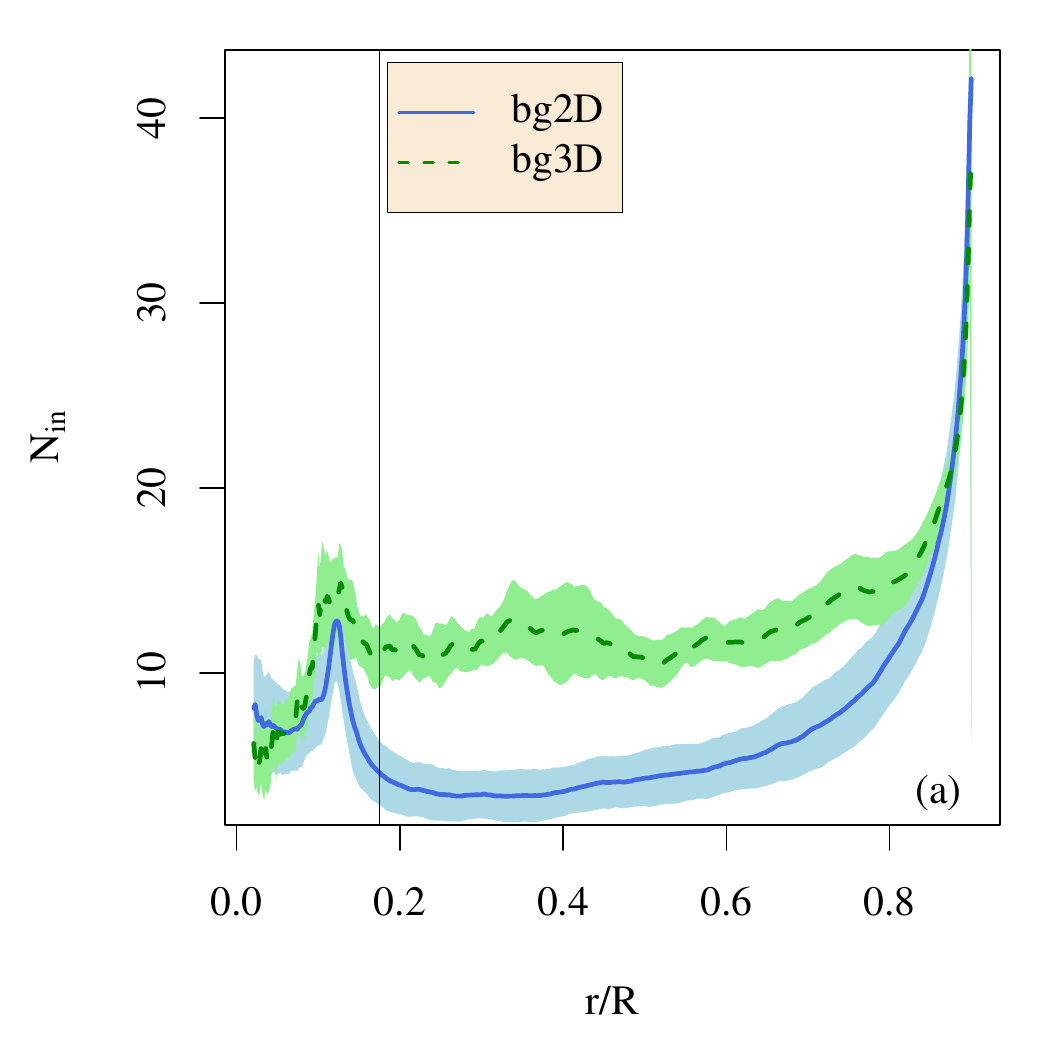}}
\resizebox{3.5in}{!}{\includegraphics{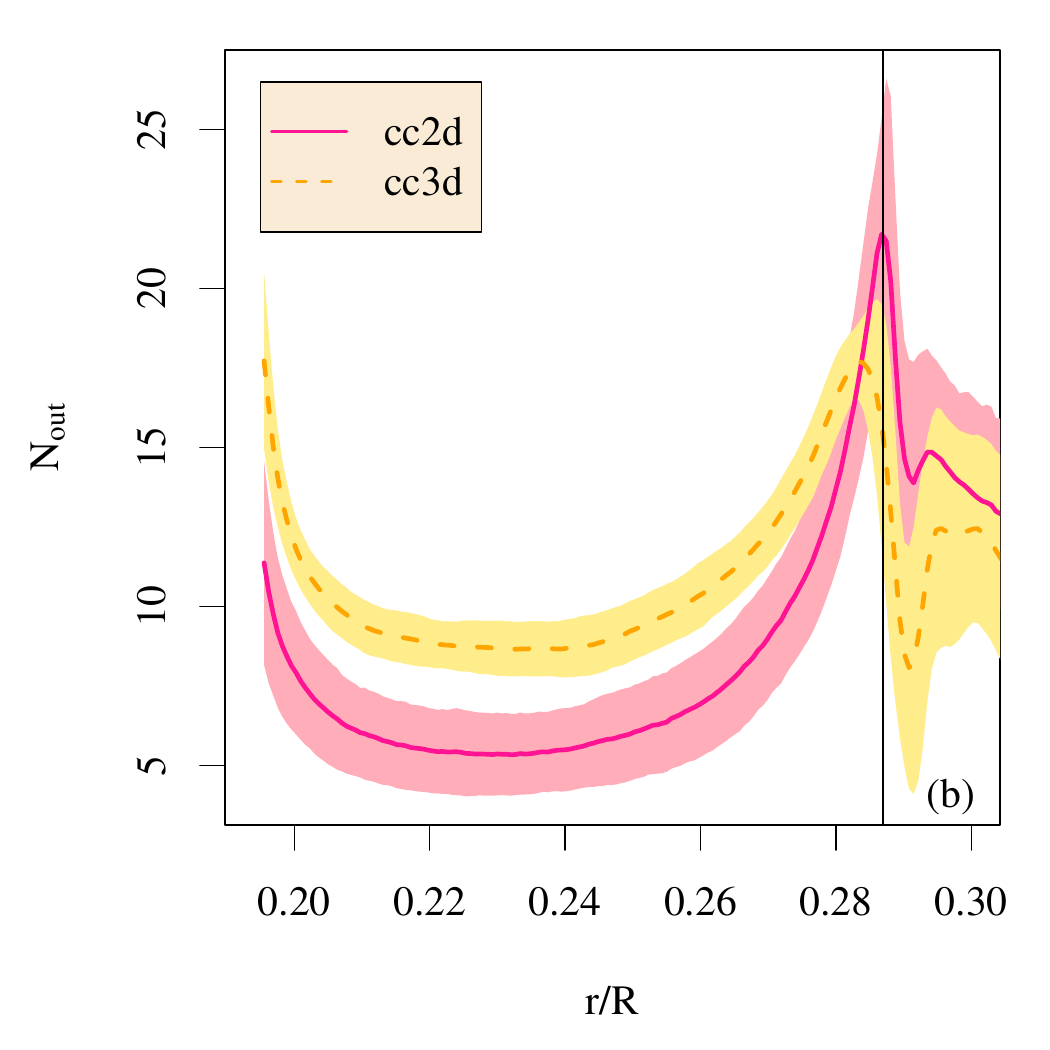}}
\caption{(a) Number of inflows vs. internal radius for 2D and 3D simulations of the $3 M_{\odot}$ red giant star, and (b) number of outflows vs. internal radius for the $20 M_{\odot}$ main-sequence star with a convective core.   The radial position of the convective boundary {determined by the Schwarzschild criterion} is indicated by a vertical black line.
\label{ninflow_plot}}
\end{center}
\end{figure}

We calculate the number of inflows, $N_{\mathsf{in}}$ by counting up the continuous regions of inflowing cells at each radius; we then average over time.  The number of outflows $N_{\mathsf{out}}$ is calculated analogously.   In each of our simulations with convective envelopes, we find the largest number of inflowing plumes at the surface.   As radius decreases, the number of plumes continually decreases until the convective boundary. The average number of inflows increases again just beyond {this} convective boundary (see Fig.~\ref{ninflow_plot}(a)), a signature that could indicate the break-up or dissipation of convective flow structures or the interaction between convection and the waves that populate the radiative region.  In the figure, the number of inflowing plumes increases rapidly for radii above $r/R= 0.7$; this could be described as a tree-like structure, with large plumes dominating at the bottom of this region and much smaller convective flow structures dominating at the top.  Below $r/R= 0.7$, the number of plumes changes only mildly until the bottom of the convection zone; this region could be described as having a ``forest-like'' structure, with convective flows of similar size dominating.
Following this analogy, a mildly ``tree-like'' structure exists in the overshooting layer, as the number of inflowing plumes increases with the depth into the overshooting layer; to our knowledge, this has not been observed before.  

In the case of core convection (see Fig.~\ref{ninflow_plot}(b)) the largest number of plumes occurs at the convective boundary. Deeper in the convection core, the number of plumes decreases and then increases toward the inner radial boundary of the simulation.  Because the convective core is comparatively small, convection appears to be entirely ``tree-like''  in this case.  This structure in the number of plumes looks highly similar to the structure that we observe in shallow outer convection zones, such as the current sun.

The picture of tree-like or forest-like convection evokes the question of whether plumes split or merge.  That question can only clearly be addressed from a Lagrangian point of view, and so it is beyond the scope of this work.

\subsection{Using plume widths and numbers in a non-dimensional parameter \label{newplumeint}}

Inspired by our relative success in using the Anders penetration parameter to differentiate between 2D and 3D simulations, we construct non-dimensional numbers from the number of plumes and their widths, using values at two radial points.  The characteristic radial profile of plume widths (see Fig.~\ref{winflow_plot}) indicates that there is a funneling effect on the plumes as they reach and then pass the convective boundary.  We therefore construct a non-dimensional parameter to indicate the strength of this funneling effect.  Because both the widths and numbers of plumes scale with the number of shear interactions between inflows and outflows, we call this the \emph{plume interaction parameter}, which we define
\begin{eqnarray}
\sigma_{\mathsf{int}} = W_{\mathsf{OL}}/W_{\mathsf{CZ}}~.
\end{eqnarray}
For each of our simulations, the plume interaction parameter is included in Table \ref{table_results}; it is always less than one for our simulations.  Resolution of the complex small-scale flows in the overshooting layer can be a challenge for diagnostics.  For simulation \emph{wide2D}, we find $\sigma_{\mathsf{int}} = 0.45$, while for simulation \emph{wm2D}, where the young sun is resolved about 4 times better, $\sigma_{\mathsf{int}} = 0.50$.  This gives a clear indication that the plume interaction parameter is sufficiently resolved in our simulations, so that it changes only a small amount with increased resolution.   The value of $\sigma_{\mathsf{int}}$ is also always larger for our 3D simulations than our 2D simulations.  This suggests that the plume interaction parameter can encapsulate a general difference between 2D and 3D stellar convection.  Using the standard deviation of the plume widths to calculate an uncertainty for the average plume interaction parameter, we find that these uncertainties are smaller than the differences between the 2D and 3D values of the plume interaction parameter in about half of our simulation pairs.  Because the plume interaction parameter is based solely on the velocity rather than the convective flux, this measure is also physically distinct from the Anders penetration parameter.

\begin{figure}
\begin{center}
\resizebox{3.5in}{!}{\includegraphics{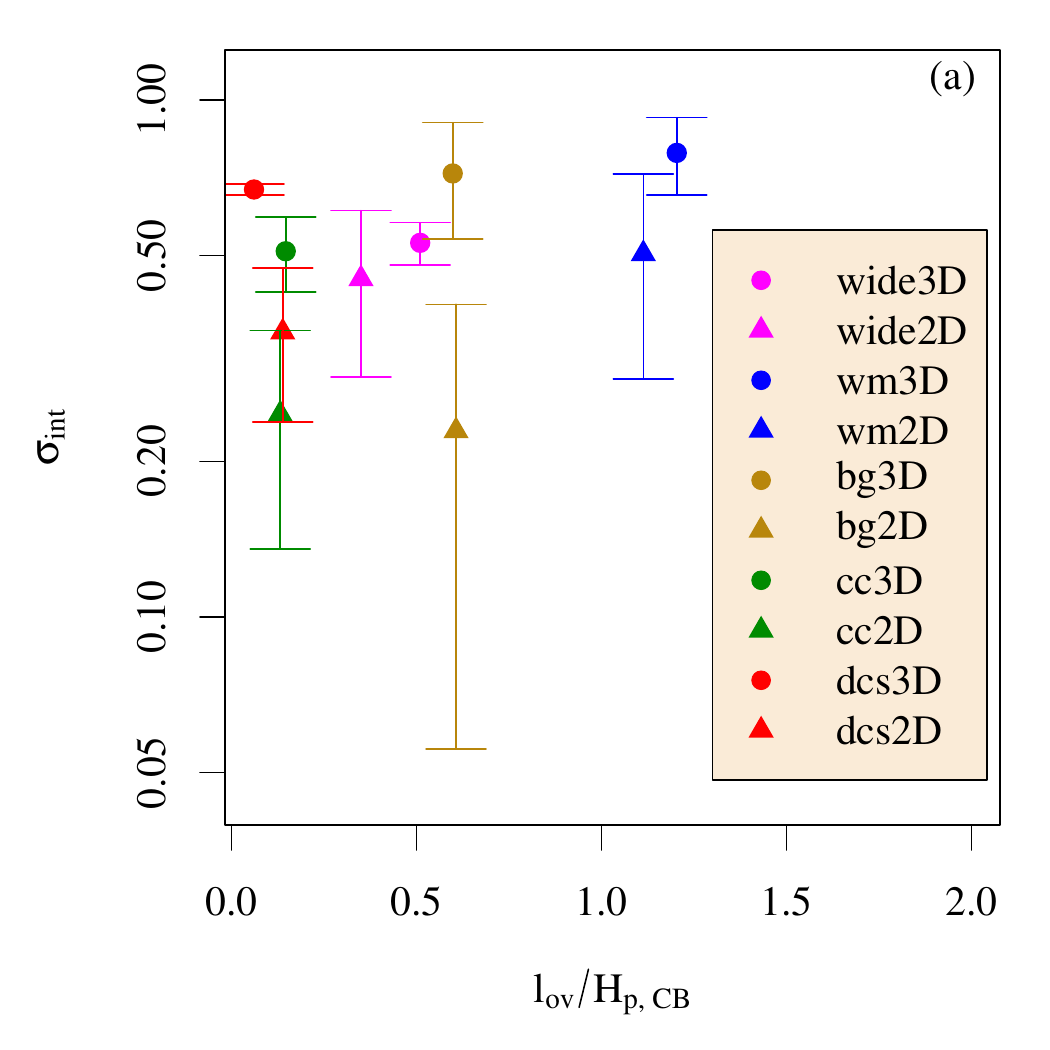}}
\resizebox{3.5in}{!}{\includegraphics{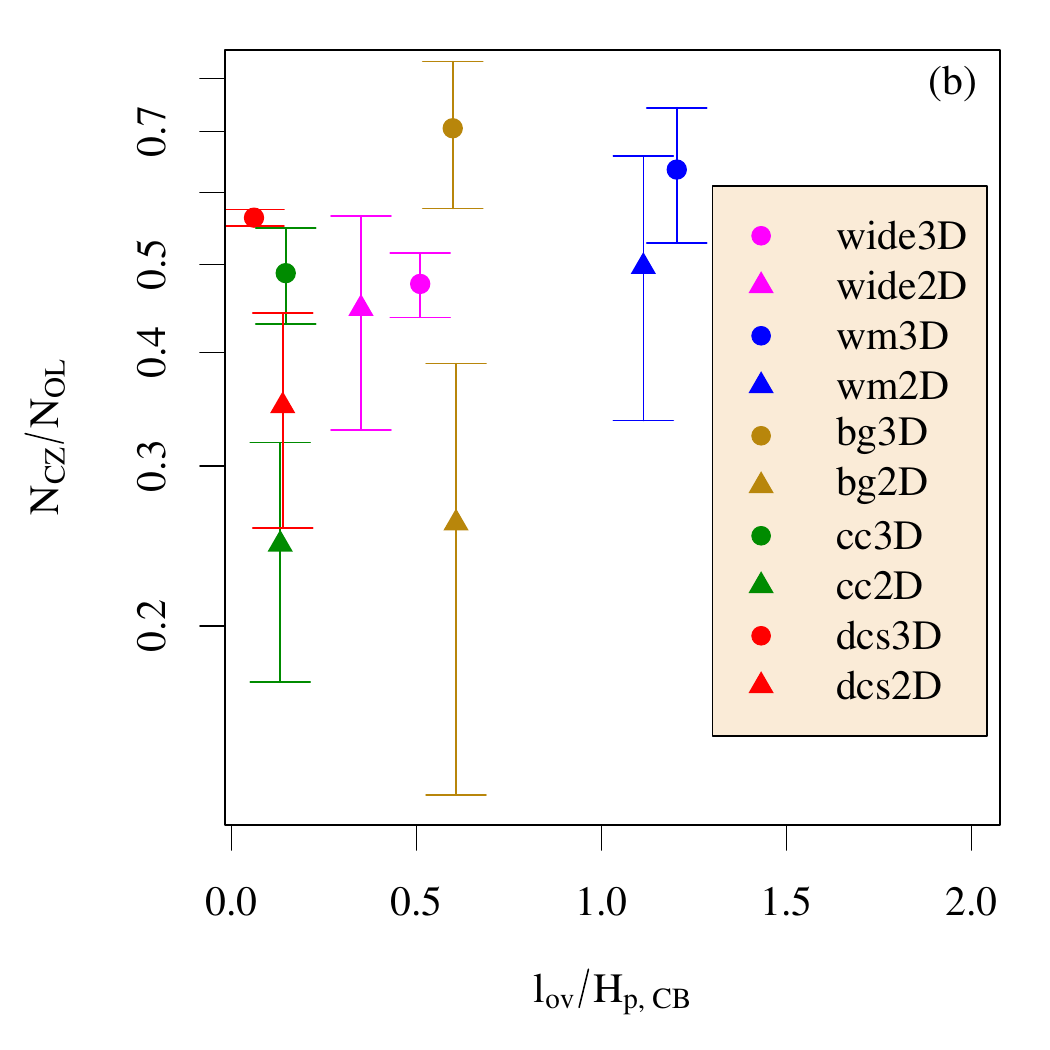}}
\caption{Nondimensional ratios (a) the plume interaction parameter, and (b) the ratio of average numbers of plumes $N_{\mathsf{CZ}}/N_{\mathsf{OL}}$ vs the overshooting depth in units of the pressure scale height at the convective boundary {determined by the Schwarzschild criterion, $\ell_{\mathsf{ov}}/H_{p,CB}$,} for all simulations studied in this work.   A logarithmic scale is applied to the vertical axis.
\label{lovvswidth}}
\end{center}
\end{figure}

In addition to the widths of plumes, the number of plumes can be combined in a ratio to produce a second nondimensional number, $N_{\mathsf{CZ}}/N_{\mathsf{OL}}$.  As with the plume interaction parameter, this number is meaningfully different for our 2D and 3D simulation pairs, with the 3D simulation always having a larger value.  Although a ratio based on the numbers of plumes does include different information from the widths of plumes, for our simulations they appear to be remarkably similar.  Because the plume interaction parameter is more clearly linked to a filling factor, we focus on that diagnostic.  Fig.~\ref{lovvswidth} shows how both of our nondimensional ratios relate to the overshooting length in our simulations.   No clear trend emerges for these pairs of simulations. However, the stellar models that we selected here are very different; investigation of a more similar set of stellar models could more easily exhibit a trend between the plume interaction parameter and the overshooting length; we are pursuing that study.

\section{Summary and Conclusions \label{secconc}}

Because 2D simulations have higher radial velocities than 3D simulations, it has generally been assumed that 2D simulations have a larger overshooting depth.  Examining convection in four different models of stars, we show that the overshooting depth is often very similar in 2D and 3D simulations.   Because 2D and 3D convective flows are visibly different, one possible explanation has been that a filling factor could explain these results.

In this work, we have studied different definitions of the filling factor for convection in realistic global simulations of stars to understand differences in 2D and 3D convection as well as their link to an overshooting depth.  Our calculations of a filling factor based on the volume percentage or based on mass flux result in characteristic profiles.  These profiles reveal that, for stars with outer convective envelopes, the inward and outward flows are highly asymmetrical near the stellar surface with a value of about one-third (for the volume-percentage filling factor) or two-thirds (for the mass-flux filling factor).  However, at the bottom of the convection zone, the filling factor is about one-half, and the convection is nearly perfectly symmetrical.  For our convective core simulation, these profiles are similar to those of a small convective envelope; at the convective boundary, there is symmetry between inflows and outflows. This is a significant new result for understanding stellar convection because it indicates that a filling factor calculated either using a volume percentage or the mass flux does not distinguish between 2D and 3D simulations.  Nor are these filling factors, or any arguments based on asymmetrical convection, able to predict the overshooting depth for different stars.

We study a filling factor based on the convective flux, as suggested by \citet{zahn1991convective}.  This calculation reveals that boundary-layer-like dynamics, as discussed by \citet{kupka2017modelling}, are significant in realistic global simulations of stars.  These flows contaminate the convective flux, distorting the signal near the convective boundary, where it might be used to predict an overshooting length.  Connected to the convective flux, we also examine the Anders penetration parameter.  This is not a filling factor because it involves two different radial points rather than a separation of inflows and outflows at a single radial point.  However, we find that this parameter can distinguish between 2D and 3D stellar convection, so it may be useful in explaining the amount of overshooting in 2D and 3D global simulations of stars, as well as for the box-in-a-star simulations where it was developed.  The convective flux in the overshooting layer is particularly intermittent, making this diagnostic converge only slowly.

We proceed to examine statistics of the widths of inward-flowing plumes (for convective envelopes) and outward-flowing plumes (for core convection).  We find clear statistical differences between 2D and 3D simulations in the standard deviation and the kurtosis.  We identify a universal shape to the radial profile of the average plume width.   We also examine the radial profile of the average number of plumes, which we link to pictures of tree-like and forest-like convection.  Based on these profiles, we observe tree-like convection in overshooting layers.  We construct a non-dimensional number from the ratio of the average radial profile of the inward plume widths at two different radial points; we call this the plume interaction parameter.  We demonstrate that the plume interaction parameter captures differences between 2D and 3D simulations.

Although both the Anders penetration parameter and our plume interaction parameter successfully and reliably indicate differences between 2D and 3D stellar convection, for the set of stars that we examine in this work, they do not clearly correlate with the overshooting length.  However, this set of stars was selected because they are very different from each other: they represent stars of different sizes, with different kinds of convection zones, at different evolutionary phases.  A more systematic study of how these parameters change with overshooting depth may successfully demonstrate their use in predicting overshooting; that work is underway.

\begin{acknowledgements}
This research is part of the Blue Waters sustained-petascale computing project, which is supported by the National Science Foundation (awards OCI-0725070 and ACI-1238993) and the state of Illinois. Blue Waters is a joint effort of the University of Illinois at Urbana-Champaign and its National Center for Supercomputing Applications.
The research leading to these results is partly supported by the ERC grants 787361-COBOM and by the STFC Consolidated
Grant ST/V000721/1.
This work used the DiRAC Complexity system, operated by the University of Leicester IT Services, which forms part of the STFC DiRAC HPC Facility (www.dirac.ac.uk). This equipment is funded by BIS National E-Infrastructure capital grant ST/K000373/1 and STFC DiRAC Operations grant ST/K0003259/1. DiRAC is part of the National E-Infrastructure.  This work also used the University of Exeter local supercomputer ISCA.  Part of this work was performed under the auspices of the U.S. Department of Energy by Lawrence Livermore National Laboratory under Contract DE-AC52-07NA27344. LLNL-JRNL-857704.
\end{acknowledgements}

\bibliographystyle{aa}
\bibpunct{(}{)}{;}{a}{}{,}
\bibliography{masscomp}

\end{document}